\documentclass[lettersize,journal]{IEEEtran}

\usepackage{amsmath,amssymb,amsfonts}
\usepackage{mathtools}
\usepackage{siunitx}

\usepackage{booktabs}
\usepackage{multirow}
\usepackage{graphicx}
\usepackage{xcolor}
\usepackage[hidelinks]{hyperref}
\usepackage{url}
\usepackage{cite}
\usepackage{acro}

\usepackage{algorithm}
\usepackage{algpseudocode}

\newcommand{\PPLedge}{\mathrm{PPL}_{\mathrm{edge}}}
\newcommand{\Pesc}{P_{\mathrm{escalate}}}
\newcommand{\Llocal}{L_{\mathrm{local}}}
\newcommand{\Lcloud}{L_{\mathrm{cloud}}}
\newcommand{\Lwan}{L_{\mathrm{wan}}}
\newcommand{\Lstall}{L_{\mathrm{stall}}}
\newcommand{\Tauroute}{\tau_{\theta}}
\newcommand{\Vpeak}{\mathcal{V}_{\mathrm{peak}}}
\newcommand{\Rmistrap}{R_{\mathrm{mis}}^{\mathrm{trap}}}
\newcommand{\Rmislat}{R_{\mathrm{mis}}^{\mathrm{lat}}}

\newcommand\blfootnote[1]{%
	\begingroup
	\renewcommand\thefootnote{}\footnote{#1}%
	\addtocounter{footnote}{-1}%
	\endgroup
}

\makeatletter
\def\footnoterule{\kern-3\p@
	\hrule \@width 3.5in \kern 2.6\p@} % the \hrule is .4pt high
\makeatother

\DeclareAcronym{ai}{ short = AI , long = Artificial Intelligence }
\DeclareAcronym{auroc}{ short = AUROC , long = Area Under the Receiver Operating Characteristic }
\DeclareAcronym{epc}{ short = EPC, long = European Patent Convention }
\DeclareAcronym{epo}{ short = EPO, long = European Patent Office }
\DeclareAcronym{fnr}{ short = FNR , long = False Negative Rate }
\DeclareAcronym{fpr}{ short = FPR , long = False Positive Rate }
\DeclareAcronym{genai}{ short = GenAI, long = Generative Artificial Intelligence }
\DeclareAcronym{ip}{ short = IP, long = Intellectual Property }
\DeclareAcronym{ipc}{ short = IPC, long = International Patent Classification }
\DeclareAcronym{llm}{ short = LLM, long = Large Language Model }
\DeclareAcronym{sme}{ short = SME, long = Small and Medium-sized Enterprise }
\DeclareAcronym{uspto}{ short = USPTO, long = United States Patent and Trademark Office }
\DeclareAcronym{oom}{ short = OOM, long = out of memory }
\DeclareAcronym{nll}{ short = NLL, long = Negative Log Likelihood }
\DeclareAcronym{gqa}{ short = GQA, long = Grouped-Query Attention }
\DeclareAcronym{wan}{ short = WAN, long = wide area network }
\DeclareAcronym{ood}{ short = OOD, long = Out-Of-Distribution }

\begin{document}

% --- Metadata --------------------------------------------------------------
\title{When Words Predict Workload}

\author{Anubhab~Banerjee \\ Nokia Germany\\ E-mail: anubhab.1.banerjee@nokia.com.}

\maketitle

\blfootnote{
This work has been submitted to the IEEE for possible publication. Personal use of this material is permitted. Permission from the author must be obtained for all other uses, in any current or future media, including reprinting/republishing this material for advertising or promotional purposes, creating new collective works, for resale or redistribution to servers or lists, or reuse of any copyrighted component of this work in other works. Copyright may be transferred without notice, after which this version may no longer be accessible.
}

%%% Abstract
\begin{abstract}
Standard distributed schedulers for \ac{llm} inference rely on static token counts or rolling latency averages, making them susceptible to failures arising from statutorily constrained text or other linguistics properties.
For example, on \ac{epo} claims, governed by the of Article~84 \ac{epc}, properties like rigidity make human and machine authorship are statistically indistinguishable. 
%Specifically, $82.8\%$ of H04L (telecom category patents) claims land in this calibrated trap band under Qwen2.5-7B (GGUF Q4\_K\_M) scoring. 
Resolving this ambiguity mid-flight forces the pipeline to dynamically expand into a heavy multi-model ensemble, triggering unpredictable KV-cache and weight-allocation spikes that saturate the VRAM ceiling of consumer-grade edge accelerators and cause severe \ac{oom} crashes and queue stalls.
To prevent this hardware collapse, we propose a CPU-side \textit{Linguistic Resource Forecasting} (LRF) gateway that extracts a 16-dimensional vector of text-structure features and processes them through an XGBoost predictor to forecast trap-band membership. 
The resulting escalation probability ($\Pesc$) is evaluated against a dynamic, closed-form routing threshold ($\Tauroute(t)$), which is recomputed per request using real-time latency telemetry. 
Crucially, the gateway safely routes requests to either the local Qwen2.5-7B edge worker or a remote Binoculars-style contrastive ensemble (Qwen2.5 7B + 32B) on an NVIDIA H100 before any edge GPU memory is unnecessarily allocated. 
In a 6,000-request live trial, the LRF gateway reduced the operational misroute fraction ($R_{\mathrm{mis}}$) to $0.087$--$0.095$---an order of magnitude below the token-count baseline ($0.849$). 
Peak edge VRAM remained safely bounded at $\SI{4.82}{\gibi\byte}$ (out of $\SI{8}{\gibi\byte}$) across a $27\times$ variation in \ac{wan} conditions. 
The XGBoost predictor achieved a live-trial AUROC of $0.84$, while the dynamic $\Tauroute(t)$ delivered an $8.2\%$ relative reduction in misroutes compared to an equivalent static threshold.
\end{abstract}


% --- Body ------------------------------------------------------------------
\section{Introduction}
\label{sec:introduction}

\IEEEPARstart{P}{roduction} serving stacks for \acp{llm} (e.g., vLLM, Triton) predominantly manage computational resources through proxy heuristics like static token counts and rolling-latency averages. 
These heuristics are predicated on the assumption that the resource footprint of an LLM request is primarily a function of its length, rather than its semantic content. 
In open-domain natural language workloads, this assumption is statistically sound, as attention mechanisms and KV-cache growth scale linearly with sequence length, making ``tokens" a reliable representation for ``workload."

However, this dependency on structural proxies creates problems in domains characterized by rigid, statutorily-constrained text registers. 
\ac{epo} patent claims, for example, are legally compelled to follow narrow, prescriptive, rigid templates, strict antecedent-basis rules, and chained subordinate clauses. 
Our previous work~\cite{ai4law2026} identified that these registers create a ``perplexity trap'', where lightweight \acp{llm} produce statistically indistinguishable likelihoods for human-authored and AI-rewritten text, causing even sophisticated detectors (e.g., Binoculars \cite{spotting}, DivScore \cite{divscore}) to collapse.

The perplexity trap is more of an imminent threat to operational stability than a failure of classification accuracy. 
In 2026-era of cascading-LLM stacks, the standard response to an ambiguous classification is to escalate the request to a heavier, cloud-based ensemble. 
Since the upstream scheduler routes based solely on token counts, it pays no attention to the claim's internal structure. 
Consequently, when a claim falls into the trap band, the gateway invokes an expensive escalation mid-flight. 
This creates a sudden, unpredictable demand on memory and latency that saturates the consumer-grade edge accelerators (e.g., NVIDIA GeForce GTX~1080), triggers \ac{oom} aborts, stalls the head-of-line queue, and forces a serial fallback to the cloud, where the system pays the full latency penalty of CUDA memory cleanup, state serialization, and remote re-transmission.

The gap in current infrastructure sits exactly on the CPU side of the gateway, \textit{before} the GPU memory is allocated. 
To close this gap, we present \textbf{Linguistic Resource Forecasting (LRF)}, a sub-\SI{5}{\milli\second} CPU pipeline that predicts hardware escalation probability $\Pesc$ using a sixteen-dimensional vector of likelihood-orthogonal features. 
By coupling this predictor with a telemetry-driven, closed-form routing threshold $\Tauroute(t)$, our system resolves operational scenarios—from VRAM-imminent overflows to WAN partitions—by overriding the escalation logic before a physical crash occurs.

\subsection{Contributions}

Our contributions in this paper are three-fold:
%\begin{enumerate}

\textbf{Linguistics-to-Hardware Telemetry.} A deterministic CPU-side mapping from sixteen static text-structure features to a downstream hardware-escalation probability $\Pesc$, computed in $O(1)$ complexity. 
This feature space combines a baseline feature with seven repurposed signals from \cite{ai4law2026} and eight new EPC-register-aware features (e.g., boilerplate frequency, POS-bigram entropy).

\textbf{Closed-Form Dynamic Routing Threshold.} A dynamic, telemetry-driven routing threshold that optimizes the balance between local edge processing and remote cloud offloading. 
This threshold is governed by an operational override hierarchy, ensuring system safety and stability in edge cases—such as network partitions or hardware overloads—where standard latency-based decision logic would be insufficient.

\textbf{Crash-Free Heterogeneous Execution.} A dual-mechanism VRAM safety interlock (closed-form pre-allocation gating + cooperative NVML-polling aborts) that maintains peak edge VRAM usage strictly within the \SI{8}{\giga\byte} ceiling on a heterogeneous GTX~1080 + H100 cluster under sustained load.
%\end{enumerate}

\subsection{Relation to Prior Work}

This paper builds partly on the \cite{ai4law2026}, which established the perplexity-trap phenomenon as a structural failure of likelihood-based AI-text detectors on \ac{epo} claims. 
While that work demonstrated the feasibility of detecting trap-band membership using a simple solution like logistic regression, the present work elevates the pipeline from offline detection to a live, distributed inference system. 
We upgrade the classifier from logistic regression to XGBoost, shift the target from semantic authorship to operational trap-band membership, and expand the topology from a single-machine bottleneck to a gateway-governed edge–cloud distributed architecture.

\subsection{Paper Organization}

The remainder of this paper is structured as follows: 
Section~\ref{sec:related-work} situates our contribution against LLM serving and dynamic routing literature. 
Section~\ref{sec:problem-statement} formalizes the perplexity-trap-to-hardware-collapse pipeline. 
Section~\ref{sec:proposed-solution} details the LRF pipeline and the derivation of our routing threshold. 
Section~\ref{sec:experiment-design} describes our heterogeneous corpus and workload harness. 
Section~\ref{sec:evaluation} reports our experimental results, and Section~\ref{sec:conclusion} concludes with a discussion of future applications.

\section{Background and Related Work}
\label{sec:related-work}

\subsection{LLM Serving and Scheduling}

Production LLM serving stacks of 2025--2026 are dominated by throughput-oriented architectures that batch independent requests and amortise GPU prefill cost. 
Representative systems include vLLM~\cite{kwon2023vllm}, NVIDIA Triton Inference Server~\cite{nvidia2024triton}, and the Orca distributed scheduler~\cite{yu2022orca}. 
Each treats every request as a length-parameterised KV-cache budget problem and routes by \textit{token count} (or a token-count-derived priority) for the edge-vs-cloud or replica-selection decision. 
The implicit assumption, that the per-token compute cost is content-invariant once $N_{\mathrm{tokens}}$ is fixed, is correct for open-domain workloads but breaks under cascading-ensemble escalation triggered by low-confidence verdicts on statutorily constrained text (this paper's central observation).

%The Llama.cpp~\cite{gerganov2023llamacpp} project is the canonical runtime for quantised LLM inference on consumer-grade Pascal-era hardware; we use it (via the \texttt{llama-cpp-python} binding) as the edge engine, configured for GGUF~Q4\_K\_M quantisation and the \texttt{sm\_61} CUDA target appropriate to a GTX~1080.

\subsection{Edge--Cloud Cooperative Inference and Dynamic Routing}

A wave of edge--cloud cooperative inference systems (speculative decoding, draft-and-verify schemes, model cascades) frames the routing decision as a per-token confidence threshold problem (e.g.\ model cascades and frugality-driven escalation rules \cite{chen2023frugalgpt}, speculative decoding with draft models~\cite{leviathan2023speculative}). 
The thresholds in these systems are typically learned offline against a fixed validation set and applied as a static cutoff at inference time. 
None of them recomputes the threshold per request from \textit{live network and queue telemetry}, and none of them recompute it from a closed-form equilibrium of expected latencies. 
The closed-form $\Tauroute(t)$ derived in \S\ref{sec:proposed-solution} fills both gaps.

\subsection{AI-Text Detection and Likelihood-Based Classifiers}

The AI-text detection literature divides into two camps: \emph{zero-shot} likelihood detectors that score a candidate text under a known scoring model and look for distributional anomalies (DetectGPT \cite{detectgpt}, Fast-DetectGPT \cite{fastdetectgpt}, Binoculars~\cite{spotting}, DivScore \cite{divscore}), and \emph{supervised} classifiers trained on labelled human/AI corpora. 
The previous paper \cite{ai4law2026} systematically benchmarked the zero-shot family on EPO patent claims and reported a structural failure mode, ``the perplexity trap", that survives detector substitution, scoring-head substitution, and LLM-family substitution. 
That paper's resolution was diagnostic (a seven-feature off-axis logistic-regression detector that recovers signal where likelihood detectors fail); the present paper recasts that diagnosis as the operational input to a distributed-systems routing problem.

The Binoculars~\cite{spotting} formulation remains our cloud-side authoritative classifier; the \cite{ai4law2026} Appendix~B result that Binoculars at Falcon-7B failed catastrophically on EPO~H04 (TP $=$ 0, FP $=$ 0, defaulting to ``Human'') is an open empirical question at the Qwen2.5 7B + 32B scale used in this paper's cloud tier (both Apache~2.0 and ungated, deliberately chosen to keep the full stack OSI-compatible open source), and the experimental design preserves the option to report either outcome (a successful escalation classifier or a failure that re-validates the hypothesis at frontier scale; see \S\ref{sec:experiment-design}).

\subsection{Linguistic Register Analysis and Stylometry}

Register-aware stylometry has a long history in computational linguistics. 
Patent-prosecution-specific work includes the M1--M6 markers identified in \cite{ai4law2026}~\S5.3 (rule-based regex anchors over EPC boilerplate phrases, claim-coverage symmetry, antecedent-basis violations). The present paper continues two of these as continuous features --- F10 (EPC boilerplate frequency, lifted from M3) and F11 (boilerplate positional variance, lifted from M5) --- and adds six new register-aware features (F3, F4, F12, F13, F14, F15) that were not in the AI4Law feature set.

Broader stylometry literature (Type-Token Ratio, Hapax-Legomena ratio, Flesch--Kincaid grade, dependency depth) supplies the remaining nine continuous features. We cite the canonical references for each feature at the point of definition in \S\ref{sec:proposed-solution}; the citation list is short because every feature has at least a four-decade history of use in computational stylometry.

\subsection{What is Missing}

The four lineages above share a structural omission. 
The LLM-serving lineage routes on token count, agnostic to text register. 
The edge--cloud-cooperative lineage uses static thresholds, agnostic to live telemetry. 
The AI-text-detection lineage frames its output as a semantic verdict, agnostic to the downstream hardware consequence. 
The stylometry lineage produces features, but does not connect them to hardware-allocation decisions.

This paper closes the omission by composing the four lineages into a single CPU-side pipeline: stylometry produces sixteen text- structure features, an XGBoost predictor maps the features to a trap-band-membership probability, a closed-form equilibrium threshold mixes that probability with live telemetry to produce a routing decision, and a two-mechanism safety interlock enforces the \SI{8}{\giga\byte} edge ceiling as a contract rather than a hope. The composition is the contribution.

%%%%%%%%%%%%%%%%%%%%%%%%%%%%%%%%%%%%%%%%%%%%%%%%%%%%
%
%
%%%%%%%%%%%%%%%%%%%%%%%%%%%%%%%%%%%%%%%%%%%%%%%%%%%%%%

\section{Problem Statement}
\label{sec:problem-statement}

\subsection{The Statutory Constraint}

Under Article~84 of the \ac{epc}, inventors are forced to use a rigid, repetitive, and narrow syntactic language for their patents. 
A claim is formally a single sentence and is anchored by mandated boilerplate phrases (``\textit{characterized in}'', ``\textit{comprising at least one}'', ``\textit{wherein the}''), enforces strict antecedent- basis between indefinite-introduction nouns (``\textit{a transceiver}'') and their definite back-references (``\textit{the transceiver}''), and chains subordinate clauses to more narrow scope. 
The result is a low-entropy text whose distributional structure is determined more by statute than by authorship.

\subsection{The ``Perplexity Trap''}

Due to the highly constrained linguistic style of patent claims, lightweight edge models struggle to distinguish between human-authored and AI-generated text based on next-token likelihood alone. 
Previous work in \cite{ai4law2026} demonstrated this limitation across various detectors, scoring heads, and generative LLMs. 
Specifically, the Weitzman overlap coefficient ($\Omega$) for per-token perplexity (PPL) distributions on the EPO H04 dataset ranges from 0.40 to 0.85. 
The most challenging generation techniques, such as standard zero-shot (Cat A) and iterative refinement (Cat D) of \cite{ai4law2026}, produced overlaps exceeding 0.75 even when scored by GPT-2-medium.

\begin{figure}[htb]
	\centering
	\includegraphics[page=1, width=0.49\textwidth]{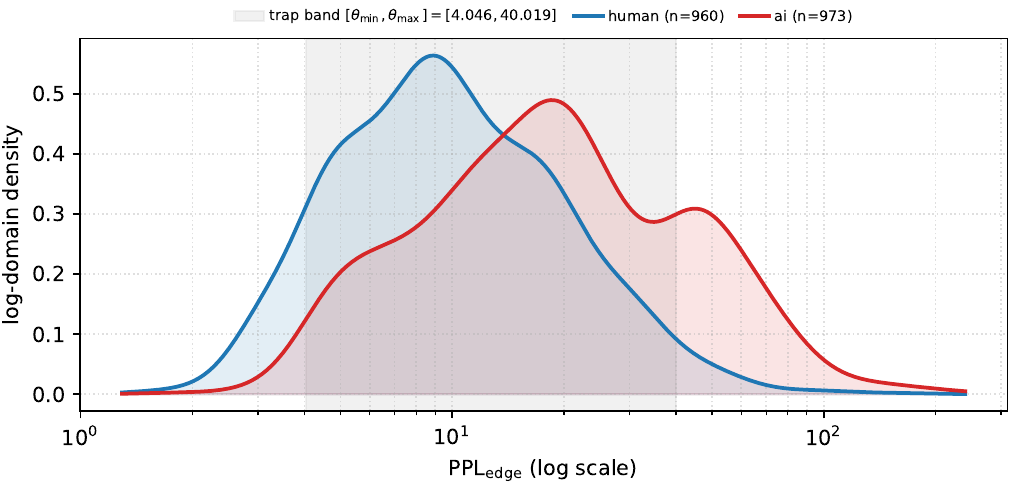}
	\vspace{-2em}\caption{Density distributions of edge-computed perplexity for human and AI-generated H04L claims}
	\label{img:prob1}
\end{figure}

Figure~\ref{img:prob1} illustrates this overlap on our deployed edge stack, using a Qwen2.5-7B-Instruct (GGUF Q4\_K\_M) model to score the H04L IID-test partition (for more details about the dataset, please refer to \S\ref{sec:experiment-design:corpus}). 
The density distributions for human and AI claims overlap almost entirely across the calibrated trap band of $[\theta_{\min}, \theta_{\max}] = [4.046, 40.019]$. 
Under offline scoring, 82.8\% of these claims (1601 out of 1933; 960 human, 973 AI) fall directly into this band, where the two classes become virtually indistinguishable. 
This is the primary reason for algorithmic failure: for the vast majority of H04L patent claims, the edge model's perplexity scores land in a region of critical ambiguity, rendering the local evaluation insufficient.

\subsection{The Dynamic Graph Expansion}

When a patent text's perplexity falls within the trap band, the edge node cannot issue a reliable decision. 
In modern cascading LLM architectures, the standard procedure for resolving this type of ambiguity is to dynamically escalate the request to a heavier, cloud-based ensemble mid-flight. 
%This expansion typically involves invoking a Binoculars formulation (e.g., a 70B observer paired with a 72B performer) \cite{spotting}, utilizing retrieval-augmented verification, or querying a supervised auxiliary classifier. 
While this dynamic escalation is the intended behavior of a cascading stack, it introduces a critical systems-level challenge: the mid-flight expansion occurs completely unannounced to the upstream scheduler.

%\subsection{The Hardware Limitation}

\subsection{The Hardware Collapse}

Mid-flight ensemble escalation produces a sudden, unpredictable spike in VRAM consumption because the system must simultaneously hold the active scoring model's KV-cache and load the newly invoked ensemble model's weights. 

Our edge node operates on a consumer-grade NVIDIA GeForce GTX 1080 (Pascal sm\_61, \SI{8}{\giga\byte} VRAM, no tensor cores). 
For the quantized Qwen2.5-7B-Instruct (Q4\_K\_M) model, the VRAM budget consists of three static components: $\approx \SI{4.6}{\giga\byte}$ of model weights, $\approx \SI{0.5}{\giga\byte}$ of \texttt{llama.cpp} runtime overhead, and the dynamic KV-cache $\mathcal{M}_{kv}(N)$. 
As detailed in \S\ref{sec:proposed-solution:safety}, we compute this KV-cache conservatively using the dense query head count ($h_q = 28$) rather than the model's actual \ac{gqa} count ($h_{kv} = 4$) to maintain strict systems safety. 
This yields a per-token footprint of $\mathcal{M}_{kv}(N) \approx \SI{0.383}{\mebi\byte} \times N$.

\begin{figure}[htb]
	\centering
	\includegraphics[width=\linewidth]{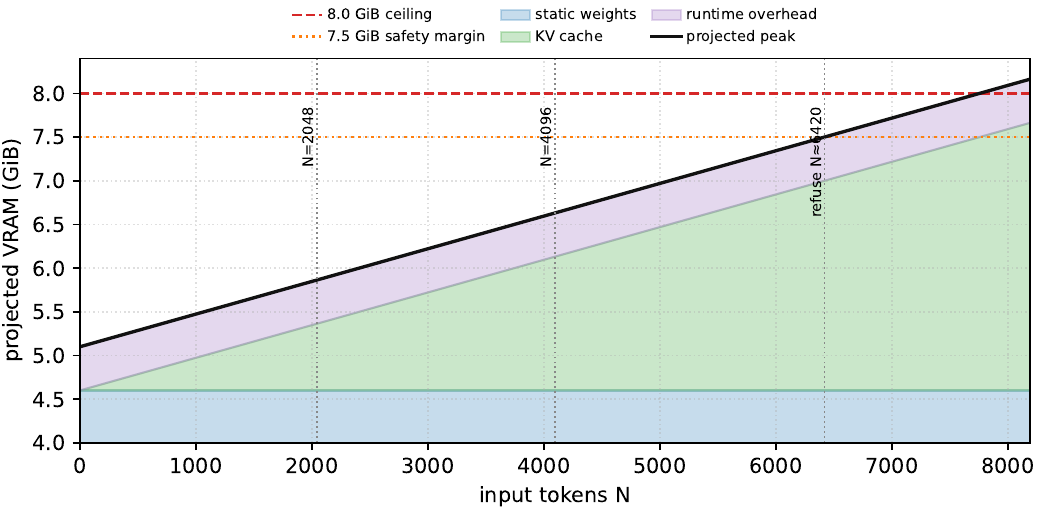}
	\vspace{-2em}\caption{Analytical VRAM envelope showing the edge GTX 1080 reaching its \SI{7.5}{\gibi\byte} safety refusal threshold at $N \approx 6420$ tokens}
	\label{img:vram_envelope}
\end{figure}

Figure~\ref{img:vram_envelope} maps this closed-form mathematical envelope. 
The projected peak VRAM reaches $\approx \SI{5.87}{\gibi\byte}$ at $N = 2048$ tokens and $\approx \SI{6.63}{\gibi\byte}$ at $N = 4096$. 
Given the physical \SI{8.0}{\gibi\byte} hardware ceiling and a \SI{7.5}{\gibi\byte} safety margin, the gateway's pre-allocation rule must refuse any request where $N \gtrsim 6420$. 
If an unannounced ensemble expansion bypasses this envelope, it immediately saturates the ceiling and trips a CUDA \ac{oom} abort. 
This failure stalls the head-of-line queue, forces a host process restart, and sequentially reissues the request to the cloud, incurring the severe $\Lstall$ penalty.

% --- FIGURE 2 ---

\begin{figure}[htb]
	\centering
	\includegraphics[width=\linewidth]{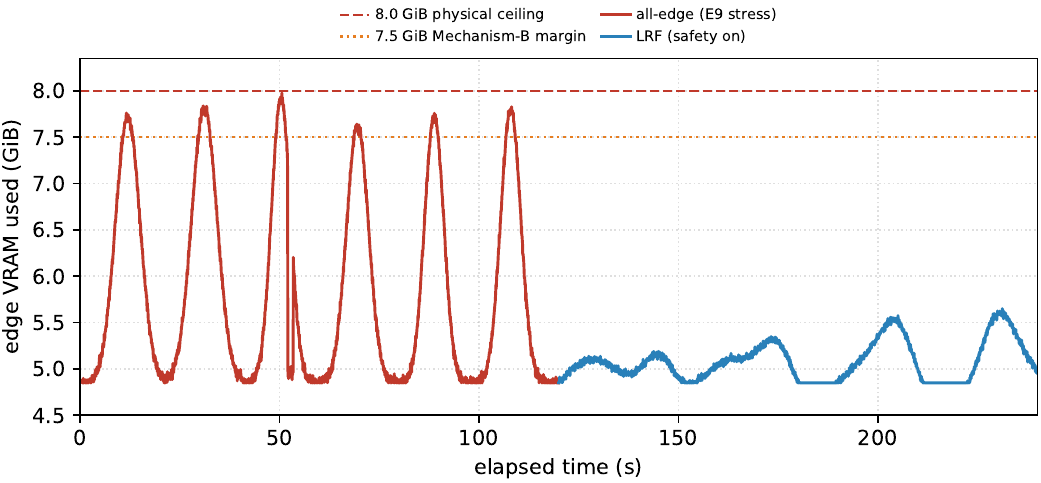}
	\vspace{-2em}\caption{Empirical edge VRAM trace under a sustained burst workload, contrasting the destructive \ac{oom} sawtooth of the all-edge policy against the stable, bounded allocations of the LRF-gated policy} 
	\label{img:vram_trajectory}
\end{figure}

Figure~\ref{img:vram_trajectory} complements this analytical envelope with high-frequency empirical NVML telemetry captured during a sustained burst stress test (Poisson arrivals at $\lambda = 4$ req/s). 
The unmitigated all-edge baseline (red trace) suffers catastrophic hardware collapse: the VRAM allocations form a destructive sawtooth pattern, repeatedly crashing into the absolute \SI{8.0}{\gibi\byte} physical ceiling and triggering hard CUDA out-of-memory aborts 
(The secondary fragmented spike near $t = 51$\,s visually captures the head-of-line queue stall, as the worker attempts to restart while the queue remains flooded). 
Conversely, when the LRF gateway and safety interlocks are engaged (blue trace), the system dynamically escalates marginal requests to the cloud, cleanly intercepting the allocation spikes and maintaining edge stability well below the \SI{7.5}{\gibi\byte} margin.

% --- FIGURE 3 ---

% This section outlines why standard routing heuristics fail for this specific workload.
% It introduces the core problem: schedulers route on length or latency, but failures depend on text structure.
\subsection{The Failure of Standard Schedulers}
\label{sec:problem-statement:scheduler-failure}

Current schedulers typically rely on token count ($N_{\mathrm{tokens}}$) or a rolling mean of recent latencies, neither of which can distinguish a standard text from a patent claim which might cause a trap-band failure. 
Trap-band membership is caused by \textit{text structure}, which is entirely different from those two metrics. 
Therefore, standard schedulers repeatedly route tasks to the edge cluster which triggers \ac{oom} aborts and degrades tail latency.

\begin{figure}[t]
	\centering
	\includegraphics[width=\columnwidth]{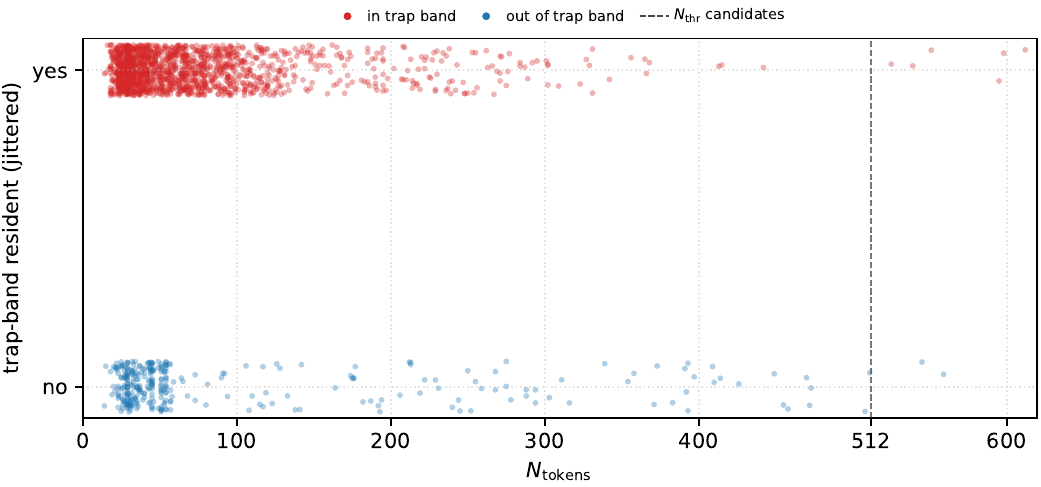}
	\vspace{-2em}\caption{Per-claim scatter of the token-count decision variable ($N_{\mathrm{tokens}}$) against the offline trap-band label.}
	\label{fig:problem-p4-tokcount-scatter}
\end{figure}

% Here we integrate the details that were previously in the Figure 4 caption.
% We explicitly mention the sample size and the jittering technique used for visual clarity.
Figure~\ref{fig:problem-p4-tokcount-scatter} plots the token-count baseline against the actual offline trap-band labels for the H04L IID-test partition ($n = 1\,933$, with the binary y-axis jittered to reveal density). 
Since H04L claims are short ($99.2\%$ fall below 512 tokens), token count proves to be an ineffective decider for this workload.  

\begin{figure}[t]
	\centering
	\includegraphics[width=\columnwidth]{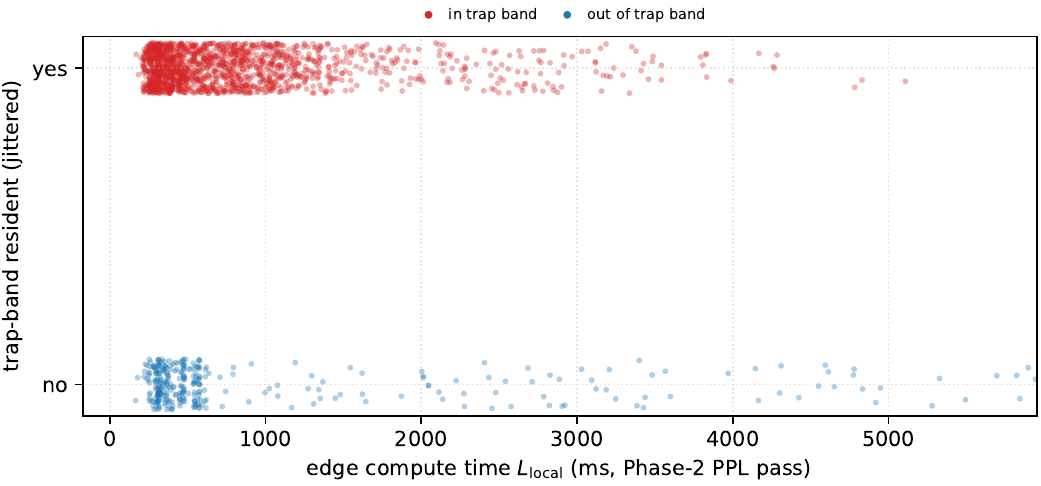}
	\vspace{-2em}\caption{Per-claim scatter of the edge compute time ($\Llocal$) against the offline trap-band label.}
	\label{fig:problem-p5-latency-scatter}
\end{figure}

% We detail the sweep grid directly in the text to compensate for the shortened caption.
Moreover, any standard routing threshold, e.g., the evaluation grid candidates $N_{\mathrm{thr}} \in \{512, 768, 1024, 1536, 2048\}$, defined in \S\ref{sec:experiment-design:baselines}, simply routes everything to edge. 
Furthermore, even at the lowest candidate threshold ($N_{\mathrm{thr}} = 512$), the in-band fraction remains stagnant at $82.8\%$, with no operating point on the $N_{\mathrm{tokens}}$ axis that separates the two classes. 
This empirical failure justifies the need for the LRF gateway evaluated in \S\ref{sec:evaluation:baseline-sweep}.
Figure~\ref{fig:problem-p5-latency-scatter} demonstrates a similar failure for the latency-mean baseline, applying the same y-axis jittering. 

% The figure environment for the token count scatter plot.
% The caption is now reduced to a single line as requested.

% We transition to discussing the latency-mean baseline, applying the same logic.
% Details from the Figure 5 caption (like jittering and the proxy metric) are absorbed here.
%We use the empirical edge compute time ($\mathrm{compute\_ms}$) from the offline Phase-2 PPL pass (\S\ref{sec:experiment-design:pilot}) as an unbiased proxy for $\Llocal$. 

% We bring the statistical medians and tails into the text body to keep the figure caption concise.
The class-conditional distributions heavily overlap. 
The median compute time for trap-band-resident claims is $\SI{596}{\milli\second}$, compared to $\SI{472}{\milli\second}$ for out-of-band claims, and both exhibit long right tails ($p_{95}$ of $\SI{2.48}{\second}$ and $\SI{4.93}{\second}$, resp.). 
Since the distributions share the same domain, any vertical decision threshold on $\Llocal$ (edge computing time) will intersect both classes, rendering it useless as a predictive routing feature.

% The figure environment for the latency scatter plot.
% The caption is strictly one line.

\subsection{The Quantitative Gap and Research Question}

Patent offices, IP law firms, and SME R\&D departments do not usually own cloud computing infrastructure of large scales, they generally use consumer-grade hardware.
Since building and maintaining such infrastructure can be expensive, it is often found that they buy or rent cloud compute time whenever necessary \cite{ai4law2026}.
This raises a big challenge: they cannot always use cloud-based deployments for every patent evaluation, otherwise it will be too expensive for them.
So, from their perspective, they need a solution, which can be used already in consumer-grade hardware, and the cloud computation will only be used when necessary.
On the other hand, the failure stages described above converge on the following:	
given a heterogeneous edge--cloud cluster, where the edge tier is a consumer-grade \SI{8}{\giga\byte} accelerator and the cloud tier is a frontier-scale H100 ensemble, can a CPU-side gateway predict and prevent hardware collapse caused by trap-band-resident requests on highly structured text, while maintaining a per-request CPU budget of $< \SI{5}{\milli\second}$, strictly respecting the \SI{8}{\giga\byte} edge VRAM ceiling, and achieving a $p_{99}$ tail latency near the theoretical network-bound minimum?

The remainder of this paper addresses this question. 
We detail the gateway architecture (\S\ref{sec:proposed-solution}), define the evaluation metrics (\S\ref{sec:experiment-design}), and report the empirical results (\S\ref{sec:evaluation}).

\paragraph*{Pilot Evidence}
A 300-row pilot has been executed as a system wiring sanity-check and is detailed in \S\ref{sec:experiment-design:pilot}. 
The conclusive answers to the research question, including the full heterogeneous-mix workload, the open-loop Poisson sweep ($\lambda \in \{0.5, 1.0, 2.0, 4.0\}$ req/s), and the baseline routing ablations, are presented in the final evaluation (\S\ref{sec:evaluation}).
\section{Proposed Solution: The LRF Gateway}
\label{sec:proposed-solution}

\subsection{System Architecture}
\label{sec:proposed-solution:architecture}

The LRF gateway sits in front of a heterogeneous two-tier cluster (Table~\ref{tab:hardware-envelope}). 
When a patent claim arrives, the gateway executes the following pipeline on a single CPU thread, with a target budget of $< \SI{5}{\milli\second}$ end-to-end:

\textbf{Pre-allocation safety check (Mechanism~B \S\ref{sec:proposed-solution:safety})}: A closed-form KV-cache size estimate confirms that the projected peak VRAM usage stays under \SI{7.5}{\giga\byte}. If not, the request is routed the request directly to cloud.

\textbf{LRF feature extraction}: Sixteen features (one systems baseline plus fifteen off-axis features) are extracted via a CPU-bound spaCy pipeline.

\textbf{XGBoost inference}: The feature vector is passed through a trained XGBoost \texttt{hist}-method booster configured for $< \SI{200}{\micro\second}$ per-row latency on a single CPU core. The booster outputs the trap-band inclusion probability $\Pesc \in [0, 1]$.
%\item \textbf{Closed-form threshold $\Tauroute(t)$.} Four telemetry metrices ($\Llocal(t)$, $\Lcloud$, $\Lwan(t)$, $\Lstall$) captured atomically at request entry feed the closed-form equilibrium equation, clamped to $[0, 1]$ and gated by an override hierarchy (VRAM-imminent $\succ$ WAN-partition $\succ$ cold-start $\succ$ closed-form), with a separate cloud-overload telemetry-substitution path that updates $\Lcloud$ upstream rather than branching the formula.

\textbf{Closed-form threshold $\tau_{route}(t)$}: At the arrival of each request, the gateway captures four live telemetry metrics: $\Llocal(t)$, $\Lcloud$, $\Lwan(t)$, and $\Lstall$. These metrics feed into a closed-form equilibrium equation to calculate the baseline routing threshold. 
The output is then normalized to a scale of $[0, 1]$. 
To ensure system safety, this baseline is preempted by critical hardware conditions according to a strict override hierarchy (VRAM-imminent $\succ$ WAN-partition $\succ$ cold-start). 
Finally, the system handles cloud overloads by substituting the $\Lcloud$ metric upstream, which safely avoids adding complex branching logic to the core equation.

\textbf{Routing decision}: If $\Pesc > \tilde\Tauroute(t)$, the request is escalated to the cloud Binoculars ensemble; otherwise it is dispatched to the local edge worker.

\begin{figure}[htb]
	\centering
	\includegraphics[page=1, width=\linewidth]{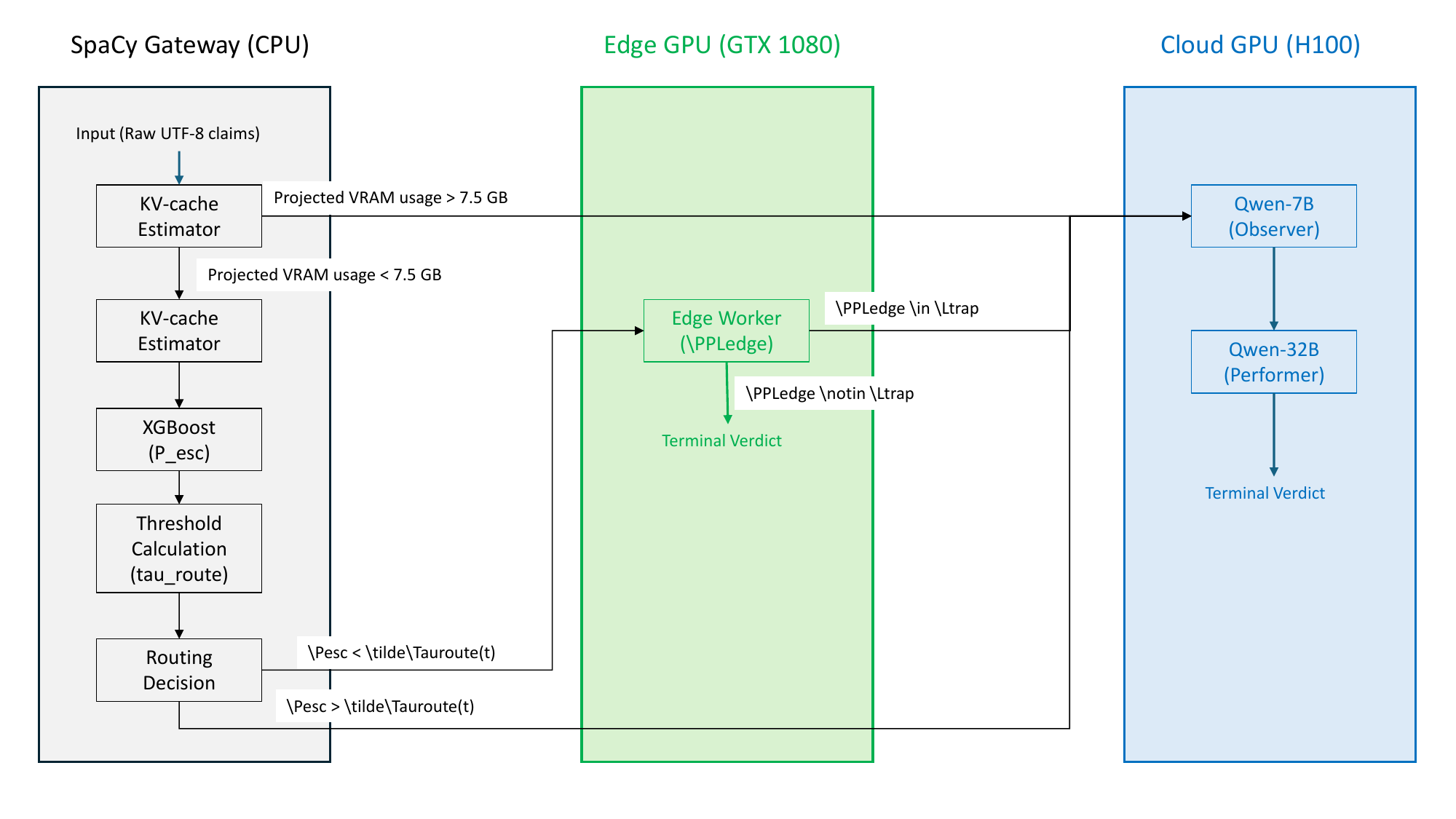}
	\caption{End-to-end system architecture diagram}
	\label{img:system_arch}
\end{figure}

The edge worker computes $\PPLedge$ on the local GTX~1080 with cooperative VRAM-abort instrumentation (Mechanism~A; \S\ref{sec:proposed-solution:safety}). 
The gateway then either returns a decision (when $\PPLedge \notin [\theta_{\min}, \theta_{\max}]$) or escalates to cloud (when $\PPLedge \in [\theta_{\min}, \theta_{\max}]$). 
This whole ensemble is depicted in Fig.~\ref{img:system_arch}.

\begin{table*}[!t]
	\caption{Hardware envelope of the heterogeneous cluster.}
	\label{tab:hardware-envelope}
	\centering
	\begin{tabular}{@{}lllll@{}}
		\toprule
		Tier & Hardware & VRAM & Architecture & Role \\
		\midrule
		Gateway & Commodity x86 CPU + RAM & --- & --- &
		LRF extraction, XGBoost inference, $\Tauroute$ recomputation \\
		Edge    & NVIDIA GeForce GTX~1080 & \SI{8}{\giga\byte} &
		Pascal sm\_61 &
		Qwen2.5-7B-Instruct GGUF Q4\_K\_M $\PPLedge$ scoring \\
		Cloud   & NVIDIA H100 (cross-region) & \SI{80}{\giga\byte} HBM3 &
		Hopper sm\_90 &
		Binoculars-style Qwen2.5 7B + 32B contrastive-PPL ensemble (Apache 2.0) \\
		\bottomrule
	\end{tabular}
\end{table*}

\subsection{The Sixteen-Feature LRF Vector}
\label{sec:proposed-solution:features}

The feature space is fixed at sixteen dimensions: one systems baseline ($N_{\mathrm{tokens}}$) and fifteen off-axis linguistic-complexity features. 
Table~\ref{tab:lrf-features} lists the vectors in index order; the order is canonical and matches the booster's stored feature-name list. 
Inherited features F1, F2, F5, F6, F7, F8, F9 lift the \cite{ai4law2026} F1--F7 set; the eight remaining off-axis features (F3, F4, F10--F15) are new for this paper.

\begin{table*}[!t]
	\caption{The sixteen LRF features (one systems baseline plus fifteen off-axis linguistic-complexity features)}
	\label{tab:lrf-features}
	\centering
	\begin{tabular}{@{}clll@{}}
		\toprule
		\# & Feature & Source & Rationale \\
		\midrule
		F0  & $N_{\mathrm{tokens}}$ &
		Tiktoken / HF fast tokenizer &
		Systems baseline; KV-cache linearity \\
		F1  & Type-Token Ratio &
		Hash set over word tokens &
		LLM-injected synonym variance vs.\ legal vocabulary reuse \\
		F2  & Hapax-Legomena Ratio &
		Word-frequency dict &
		AI4Law's strongest discriminator ($|d|=1.17$) \\
		F3  & Functional Stop-Word Entropy &
		Shannon entropy over 50 stop-words &
		LLM-specific stop-word distribution skew \\
		F4  & POS Bigram Transition Entropy &
		spaCy POS tagger + bigram entropy &
		Patent claim syntactic predictability \\
		F5  & Mean Dependency Depth &
		spaCy parser, root-to-leaf BFS &
		Heavily-nested claim drafting depth \\
		F6  & Subordinate-Clause Ratio &
		spaCy dependency tags &
		Scope-narrowing clause chains \\
		F7  & Noun-Phrase Density &
		spaCy noun chunks / words &
		Patent components as noun phrases \\
		F8  & Flesch--Kincaid Grade &
		Syllable counter + Flesch formula &
		LLM-driven jargon inflation \\
		F9  & Sentence-Length Variance &
		Per-sentence word-count variance &
		LLM-style rhythmic cadence vs.\ legal uniformity \\
		F10 & EPC Boilerplate Frequency &
		Regex over top-20 EPO/USPTO anchors &
		Mandated anchor under-utilisation by LLMs \\
		F11 & Boilerplate Positional Variance &
		Std.\ dev.\ of anchor token-index &
		Anchor-placement law violation by LLMs \\
		F12 & Semicolon-to-Word Ratio &
		Character count &
		LLM normalisation of legal punctuation cadence \\
		F13 & Inter-Delimiter Length Variance &
		Variance of semicolon-chunk lengths &
		Uneven LLM-generated claim-element lists \\
		F14 & Definite/Indefinite Noun Ratio &
		Article-tag counter &
		Antecedent-basis drift in LLM rewrites \\
		F15 & Lexical Overlap Across Claims &
		Jaccard between first 25\,\% / last 75\,\% &
		Independent-claim vocabulary continuity \\
		\bottomrule
	\end{tabular}
\end{table*}

The extractor is single-threaded and stateless across requests. 
spaCy is configured to load the \texttt{en\_core\_web\_sm} pipeline once at gateway boot, with NER, lemmatiser, and attribute-ruler disabled Configuration to meet the criteria of being lightweight with strict latency requirement. 
A pilot extraction over a 300-row sample of H04L IID claims completes at $4.6$~ms per row on the development host (a laptop CPU also feeding the GPU), while the deployment-grade gateway with a dedicated CPU core targets the documented $< \SI{5}{\milli\second}$ envelope.

% ============================================================================

\subsection{XGBoost Predictor}
\label{sec:proposed-solution:predictor}

The predictor simply follows this equation:
\begin{equation}
	y(c) =
	\begin{cases}
		1 & \text{if } \PPLedge(c) \in [\theta_{\min}, \theta_{\max}] \\
		0 & \text{otherwise}
	\end{cases}
\end{equation}

The predictor estimates trap-band membership of the edge-PPL for the deployed scoring model, not for human-vs-AI authorship. 
Under the operational target, a ``false positive'' is a wasted cloud escalation, not a false accusation of an author. 
Authorship classification is delivered downstream, either terminally at the edge (when $\PPLedge \notin [\theta_{\min}, \theta_{\max}]$) or at the cloud (when the Binoculars ensemble is invoked).

\paragraph*{Model configuration.}
The XGBoost predictor is configured with a binary objective and a CPU-bound histogram tree method (\texttt{hist}). 
Model hyperparameters include a maximum depth of 6, a learning rate of 0.05, and a 0.8 ratio for both subsampling and column sampling. 
The model trains for a maximum of 500 boosting rounds, utilizing early stopping with a patience of 20 rounds on validation log-loss. 
To strictly prevent data leakage, training uses patents from five different patent families. 
During inference, the gateway calls \texttt{Booster.inplace\_predict} on a pre-allocated NumPy buffer rather than constructing a new \texttt{DMatrix} per request; this optimization is critical for keeping single-row prediction latency below the \SI{200}{\micro\second} budget.

% ============================================================================

\subsection{Edge Perplexity Computation $\PPLedge$}
\label{sec:proposed-solution:edge-ppl}

For an input claim tokenised into $X = (x_1, \dots, x_N)$ under the edge model's vocabulary, the edge perplexity is the exponentiated mean negative log-likelihood:

\begin{equation}
	\PPLedge(X) = \exp\!\left(-\frac{1}{N}\sum_{i=1}^{N} \log P_\theta(x_i \mid x_{<i})\right).
\end{equation}

We adopt the following five-step pipeline: 
(i) sub-word tokenisation, 
(ii) autoregressive forward pass,  
(iii) log-softmax in log-domain, %(never round-tripping through linear-domain probabilities, which underflow on any logit gap larger than 88 on float32 / 745 on float64), 
(iv) \ac{nll} accumulation in float64 and 
(v) scalar exponentiation. 
Inputs longer than $n_{\mathrm{ctx}} = 4096$ tokens are chunked with a 256-token rolling overlap, with per-chunk perplexities aggregated as a token-count-weighted geometric mean.

The edge engine is a \texttt{llama-cpp-python} wrapping the llama.cpp CUDA backend compiled for Pascal sm\_61. 
At inference time the engine offloads all transformer layers to the GTX~1080 (\texttt{n\_gpu\_layers $= -1$}). 
The wrapper exposes a cooperative \texttt{threading.Event} abort flag that the inference loop checks between tokens; setting the flag from the VRAM watcher thread raises \texttt{EdgeAbortError} at the next token boundary (see \S\ref{sec:proposed-solution:safety}).

\paragraph*{Pilot measurement.}
A 300-row class-balanced stratified sample of the H04L IID test partition (\S\ref{sec:experiment-design}) was scored at \SI{4.7}{\minute} of wall-clock time on the GTX~1080 after the CUDA-enabled rebuild of \texttt{llama-cpp-python} (a CPU-only build of the same code completed the same claim at $\approx 5$~tok/s versus $\approx 86$~tok/s on the GPU, a $\sim 17{\times}$ speedup). 
Class-conditional summary statistics are reported in \S\ref{sec:experiment-design:pilot}.

% ============================================================================

\subsection{Weitzman $\alpha$-Mass Trap-Band Calibration}
\label{sec:proposed-solution:trap-band}

The perplexity trap band $[\theta_{\min}, \theta_{\max}]$ is derived offline through a five-step calibration procedure:
%\begin{enumerate}

\textbf{Perplexity Computation:} We compute the local perplexity $\PPLedge$ for every text sample in the training corpus (H04L samples consist of patent claims only, whereas G06F and A61K samples have patent abstracts to create a heterogeneous evaluation corpus).

\textbf{Density Estimation:} We apply Gaussian kernel density estimation to model the probability distributions of the log-perplexities for both human-authored and AI-generated texts.

\textbf{Overlap Calculation:} We compute the Weitzman overlap coefficient ($\Omega$) by determining the intersection area of these two density curves.

\textbf{Band Extraction:} We identify the narrowest contiguous log-perplexity interval that captures 90\% ($\alpha = 0.90$) of this total overlap area.

\textbf{Linear Conversion:} The interval boundaries are scaled back to the linear domain to define the final thresholds $\theta_{\min}$ and $\theta_{\max}$ for request routing.
%\end{enumerate}

A pilot run on a 300-sample subset yielded an overlap coefficient of $\Omega = 0.756$, resulting in a calibrated trap band of $\theta_{\min} = 4.53$ and $\theta_{\max} = 47.74$. 
The at-scale deployment follows this exact procedure across the complete corpus.

% ============================================================================
\subsection{Closed-Form Dynamic Routing Threshold $\Tauroute(t)$}
\label{sec:proposed-solution:tau}

The routing threshold represents the closed-form equilibrium between two expected-latency equations under a Bernoulli escalation model:

\begin{align}
	E[L_{\mathrm{edge}}]
	&= (1 - \Pesc)\Llocal(t) + \Pesc \cdot \nonumber \\
	&\quad \bigl(\Llocal(t) + \Lstall + \Lwan(t) + \Lcloud\bigr), \\
	E[L_{\mathrm{cloud}}]
	&= \Lwan(t) + \Lcloud.
\end{align}

Equating these expectations and solving for the equilibrium escalation probability $\Tauroute(t)$ yields, after the $\Llocal$ terms cancel:

\begin{equation}
	\Tauroute(t) = \frac{\Lwan(t) + \Lcloud - \Llocal(t)}
	{\Lstall + \Lwan(t) + \Lcloud}.
	\label{eq:tau-theta}
\end{equation}

The operational threshold is normalized to $\tilde\Tauroute(t) = \min(1, \max(0, \Tauroute(t)))$. 
The router escalates the request to the cloud iff $\Pesc > \tilde\Tauroute(t)$.

\subsubsection{Operational Control Scenarios}
Equation~(\ref{eq:tau-theta}) remains valid only where all four telemetry quantities are finite and the cluster is stable. 
To maintain system safety across eight distinct operational scenarios (cold start, steady-state light queue, steady-state heavy queue, degraded WAN, WAN partition, VRAM-imminent override, cloud overload and joint stress), we implement a two-class override architecture.

The first class consists of \textit{short-circuit} overrides that completely bypass the closed-form equation, applying instead a deterministic constant for $\tilde\Tauroute(t)$. 
These follow a strict priority hierarchy: VRAM-imminent ($\tilde\Tauroute = 0$) $\succ$ WAN-partition ($\tilde\Tauroute = 1 - 10^{-3}$) $\succ$ cold-start ($\Tauroute^{(0)} = 0.50$) $\succ$ standard closed-form evaluation.

The second class is a \textit{telemetry-substitution} override designed specifically to handle cloud overloads which sits outside the standard evaluation chain. 
Through passive RTT-decomposition, the gateway substitutes the live windowed-mean $\Lcloud$ for its nominal static value upstream. 
As a result, the closed-form formula dynamically adapts its evaluation using the degraded $\Lcloud$ metric.

\subsubsection{Per-request Update Cost}
Recomputing $\Tauroute(t)$ incurs minimal computational overhead: four memory reads, one subtraction, two additions, one division, and one normalization.
Empirically, this completes in $< \SI{100}{\micro\second}$ on a commodity x86 CPU core, fitting comfortably within the gateway's overarching $< \SI{5}{\milli\second}$ latency budget.

% ============================================================================
\subsection{Hybrid VRAM Safety Interlock}
\label{sec:proposed-solution:safety}

% The introduction establishes the dual-mechanism approach to enforcing the hardware limit.
The \SI{8}{\giga\byte} edge VRAM ceiling is strictly enforced by two cooperating mechanisms: a runtime polling watcher (Mechanism A) and a pre-allocation admission gate (Mechanism B).

\paragraph*{Mechanism A (Cooperative NVML polling)}:
% Details the daemon thread and its low-overhead polling configuration.
To capture unpredictable memory spikes during execution, a background daemon thread runs continuously on the edge worker. 
Operating at a \SI{50}{\milli\second} cadence with a negligible \SI{0.2}{\percent} CPU idle cost, the daemon polls the GPU memory state via NVML. 
% Explains the abort trigger and the cooperative nature of the stop.
If the utilized VRAM exceeds \SI{95}{\percent}, the daemon atomically sets a shared abort flag.
Since driver-level preemption is unavailable on the Pascal sm\_61 architecture, the \texttt{llama.cpp} inference loop cooperatively checks this flag at every token boundary, raising an abort exception if tripped. 
% Details the hysteresis loop and the routing override link.
To prevent rapid oscillation during the post-abort memory drain, the flag relies on a hysteretic clear threshold of \SI{85}{\percent}. 
Each abort logs the full causal chain (completed tokens, VRAM state, subsequent route, and observed stall). 
Crucially, tripping this flag also broadcasts a \textit{VRAM-imminent override} to the upstream $\Tauroute(t)$ pipeline, forcing all new requests to the cloud until the local edge VRAM falls back below the \SI{85}{\percent} safe threshold.

\paragraph*{Mechanism B (Pre-allocation closed-form check)}:
% Explains the mathematical prediction model for KV-cache size.
Operating at the gateway prior to admission, Mechanism B computes the projected peak VRAM using the formula $\mathrm{static\_weights} + \mathcal{M}_{kv}(N) + \mathrm{runtime\_overhead}$. 
The KV-cache size $\mathcal{M}_{kv}(N)$ is calculated as $2 \cdot B \cdot N \cdot l_{\mathrm{layers}} \cdot h_{kv} \cdot d_{\mathrm{head}} \cdot \mathrm{BPP}$. 
% Details the specific model parameters and the critical GQA design choice.
For the Qwen2.5-7B architecture, this uses $l = 28$, $d = 128$, batch size $B = 1$, and $\mathrm{BPP} = 2$ (representing fp16 precision for keys and values in \texttt{llama.cpp}). 
Although Qwen2.5-7B utilizes \ac{gqa} with $h_{kv} = 4$, our safety check intentionally applies the dense baseline of $h_{kv} = h_{q} = 28$. 
This conservative approach provides headroom to absorb second-order memory bloat caused by \texttt{llama.cpp}'s internal KV layout, \texttt{cudaMalloc} fragmentation, and runtime CUDA kernel scratch space on Pascal GPUs. 
% Translates the formula into hard byte estimates and the final decision rule.
This non-\ac{gqa} constant yields a per-token KV footprint of approximately \SI{0.383}{\mebi\byte} $\times N$, projecting a peak of $\approx \SI{5.9}{\gibi\byte}$ at $N = 2048$ and $\approx \SI{6.6}{\gibi\byte}$ at $N = 4096$. 
The router preemptively escalates any request to the cloud if this projected peak, plus a \SI{0.5}{\gibi\byte} safety margin, exceeds the \SI{8}{\gibi\byte} ceiling (which occurs at $N \approx 6\,420$ tokens). 
% Provides the empirical proof that the mechanism works without false positives.
During live trials, the measured peak was $\Vpeak \approx \SI{4.82}{\gibi\byte}$, confirming that while conservative, the bound does not trigger spurious cloud escalations.

% Concluding summary tying both mechanisms together.
While Mechanism~A acts as a runtime fail-safe against hardware-level memory fragmentation and overhead drift, Mechanism~B intercepts mathematically unfeasible long claims at the gateway before they can consume bandwidth.

%==============================================================================
% Section V — Experiment Design.
%
% Story: the metric framework (\S V.A), the heterogeneous-mix corpus
% (\S V.B), the hardware envelope and workload (\S V.C), the baselines
% (\S V.D), the ablation schedule (\S V.E), and finally a snapshot of
% the pilot run that has already executed (\S V.F).
%
% The pilot subsection is the one place where measured numbers appear
% in this draft. Every number is from artifacts/ on disk; no number
% is invented.
%==============================================================================

\section{Experimental Methodology}
\label{sec:experiment-design}

\subsection{Metric Framework}
\label{sec:experiment-design:metrics}
To evaluate the system, we measure seven performance indicators categorized into three distinct domains as below.

\subsubsection{ML Predictor Metrics (Gateway XGBoost)}

\textbf{AUROC:} The Area Under the Receiver Operating Characteristic curve for the XGBoost predictor. 
This tracks the model's accuracy in identifying claims that fall into the "trap band" (defined in \S\ref{sec:proposed-solution:predictor}). 
Our performance target (operational green-zone) is an AUROC $\geq 0.85$.
This threshold acts as the operational gating requirement, ensuring the predictor possesses sufficient discriminatory power to balance edge-collapse avoidance (false negatives) against unnecessary WAN utilization (false positives).

\textbf{\ac{fpr} at TPR $\geq 0.80$:} The fraction of "safe" claims that the gateway incorrectly escalates to the cloud when the system is tuned to capture at least 80\% of true trap-band claims, quantifying the frequency of unnecessary escalations. 
Our performance target is an FPR $< 0.25$.
This limit ensures that while the gateway aggressively intercepts at least 80\% of hardware-threatening claims to maintain edge stability, it does not needlessly degrade into an all-cloud routing policy by wasting expensive H100 compute and WAN bandwidth on more than a quarter of safely edgeable requests.

\subsubsection{Distributed-Systems Metrics}

\textbf{$p_{99}$ tail latency:} The end-to-end RPC turnaround time, captured across the primary H04L trial ($n = 6\,000$ requests \S\ref{sec:experiment-design:operating-point}) and remeasured per arrival-rate cell in the auxiliary $\lambda$-sweep harness ($4 \times 5\,000 \times 3 = 60\,000$ open-loop Poisson samples). 
Performance targets: $\leq \SI{500}{\milli\second}$ is the goal, whereas $\geq \SI{2000}{\milli\second}$ marks a legacy failure mode (see \S\ref{sec:evaluation:static-vs-dynamic} for the target gating Table~\ref{tbl:static-vs-dynamic}).

\textbf{Primary misrouting rate ($\Rmistrap$):} Calculated as $(\mathrm{LRF\ FP} + \mathrm{LRF\ FN})/N$, evaluated against the offline-measured trap-band membership of $\PPLedge$ as the ground truth. Performance target: $< 0.15$.

\textbf{Secondary misrouting rate ($\Rmislat$):} The fraction of requests where an offline oracle confirms the unchosen routing path would have been strictly faster. 
This is reported for analytical context without a binary pass/fail threshold.

\textbf{Gateway extraction overhead:} The total end-to-end CPU time required for LRF feature extraction, XGBoost inference, and $\Tauroute(t)$ recomputation. Performance target: $< \SI{5}{\milli\second}$ per request.

These targets act as practical engineering goals to govern system efficiency, providing baseline pass/fail criteria for our subsequent live-trial evaluations rather than universally mandated thresholds.

\subsubsection{Hardware-Constraint Metrics.}

\textbf{Peak edge VRAM ($\Vpeak$):} The maximum NVML-polled \texttt{used\_bytes} on the GTX~1080 during the 6,000-request primary trial. (Note: The $\lambda$-sweep harness reports its own per-cell $\Vpeak$, but Table~\ref{tbl:static-vs-dynamic} utilizes the claim-level replay data). Performance targets: $\Vpeak < \SI{7.5}{\gibi\byte}$ ensures operational safety, while $\Vpeak \geq \SI{8.0}{\gibi\byte}$ triggers a hard \ac{oom} failure.

\textbf{Cross-cluster \ac{wan} bandwidth:} The total bytes transmitted from the gateway to the cloud over a sustained one-hour run. 
Performance targets: $< \SI{50}{\mega\byte\per\hour}$ indicates an effective filter, whereas $> \SI{500}{\mega\byte\per\hour}$ suggests excessive cloud escalation.

%================

\subsection{Corpus: Heterogeneous-Mix Design}
\label{sec:experiment-design:corpus}

The dataset has a heterogeneous-mix to evaluate the predictor's robustness across different \acp{ipc}, generator families, and prompt categories. We utilize two data generation recipes:
\begin{itemize}
	\item \textbf{Recipe A (Mixed-Generator):} Categories A--E (from \cite{ai4law2026}) utilize a balanced $50/50$ split of AI rewrites generated by Claude Opus 4.6 and Qwen 2.5-72B-Instruct.
	\item \textbf{Recipe B (Cross-Generator Probing):} Categories A--E are exclusively generated by Claude Opus 4.6. A held-out Category F is generated entirely by Qwen 2.5-3B-Instruct (a model family the classifier never observes during training). This recipe explicitly tests the predictor's categorical cross-generator robustness.
\end{itemize}

As detailed in Table~\ref{tab:dataset}, the four-\ac{ipc} main partition and the one-IPC long-tail diagnostic collectively yield 4,100 abstracts and 25,138 claims. 
AI-generated instances carry a category-letter suffix (e.g., \texttt{EP3533259B1F}), ensuring deterministic traceability back to the human one. 
The dataset is split into training (70\%), validation (15\%), and IID-test (15\%) partitions at the patent-family level using a fixed seed (seed = 42). 
The C07D IPC is strictly held out as an \ac{ood} test set, and F03D serves as a sparse-class diagnostic.

\begin{table*}[!t]
	\caption{Heterogeneous-mix corpus, pairing symmetry is enforced at the abstract level across all IPCs}
	\label{tab:dataset}
	\centering
	\begin{tabular}{@{}llllrr@{}}
		\toprule
		Tier & IPC & Recipe & Notes & Humans & AI rewrites \\
		\midrule
		IID       & H04L & B         & Verbatim mixed + robustness pools   & 500 abstracts + 6,613 claims & 600 abstracts + 5,923 claims \\
		IID       & G06F & A         & Paired Claude+Qwen-72B              & 500 abstracts + 6,186 claims & 500 abstracts + 0 claims \\
		IID       & A61K & A-partial & Claude-only                         & 280 abstracts + 0 claims      & 280 abstracts + 0 claims \\
		OOD       & C07D & A++       & Claude-oversampled, mixed OOD       & 711 abstracts + 6,416 claims & 711 abstracts + 0 claims \\
		long-tail & F03D & minimal   & Sparse-class diagnostic             & 9 abstracts + 0 claims        & 9 abstracts + 0 claims \\
		\bottomrule
	\end{tabular}
\end{table*}

\subsection{Hardware Envelope and Workload}
\label{sec:experiment-design:hardware}

The deployment cluster operates across a heterogeneous architecture. 
The gateway and edge worker are co-located on-premises in an EU-Central office premise (NVIDIA GeForce GTX 1080). 
The cloud worker operates on a single NVIDIA H100 80GB in a separate EU region, accessed via gRPC and TLS 1.3 over HTTP/2, utilizing gzip compression for payloads exceeding 4 KB.

\paragraph*{Cloud Kernel Selection}
The cloud-side Qwen2.5 7B + 32B AWQ ensemble is executed using the pure-PyTorch \texttt{AWQ\_TORCH} reference kernel, avoiding the Marlin (sm\_80+) JIT kernel to ensure strict environmental reproducibility. 
This deliberately inflates the cloud compute envelope ($\Lcloud$) by a factor of $2\times$--$3\times$ ($\approx 80$~ms versus $\approx 30$~ms per claim on the 32B model). 
This conservative configuration ensures our evaluation is strictly pessimistic: by artificially inflating the penalty of every cloud escalation, the reported $p_{99}$ tail latency represents a possible worst-case scenario.

\paragraph*{Workload Drivers}
We evaluate the system using two distinct workload drivers:

\textbf{Arrival-Rate Variance Harness:} Emits open-loop Poisson arrivals with inter-arrival times sampled from $\mathrm{Exp}(\lambda)$ across $\lambda \in \{0.5, 1.0, 2.0, 4.0\}$ req/s. 
We generate 5,000 synthetic events per $\lambda$ point across three variance seeds (42, 43, 44), totaling 60,000 RPCs per router configuration. 
This harness tests system stability under varying loads.

\textbf{Headline Claim-Level Replay:} A deterministic, closed-loop replay of the de-duplicated H04L claim pool. It evaluates 1,500 claims across four seeds (42, 43, 44, 45) for a total of 6,000 routed requests, completely isolating the threshold policy's performance from arrival-distribution variance.

\paragraph*{Telemetry}
The dynamic routing threshold $\Tauroute(t)$ consumes four atomically captured telemetry metrics: $\Llocal(t)$ via a queue-depth-scaled EWMA of recent edge completions; $\Lwan(t)$ from 1~Hz active gRPC probes fused with passive RTT decomposition; $\Lcloud$ from the cloud's self-reported compute time; and $\Lstall$ derived from a forced CUDA OOM fault-injection test.

\paragraph*{Operating Point for the Headline Live Trial}
\label{sec:experiment-design:operating-point}

The closed-form $\Tauroute(t)$ supports two execution modes. 
To isolate the XGBoost predictor's raw efficacy from the dynamic controller's adjustments, the 6,000-request headline trial utilizes a frozen-seed mode. 
The four latency arguments are pinned to static estimates: $\Llocal = 300$~ms, $\Lwan = 80$~ms, $\Lcloud = 1800$~ms, and $\Lstall = 2000$~ms. 
This yields a static operating equilibrium:
\[ \Tauroute^{\star} \;=\; \frac{\Lwan + \Lcloud - \Llocal}{\Lstall + \Lwan + \Lcloud} \;=\; \frac{1580}{3880} \;=\; 0.407 \]
This isolates $P_{\mathrm{esc}}$ performance, allowing the AUROC and FPR@TPR$\geq 0.80$ metrics to reflect solely the classifier's accuracy at a fixed decision boundary.

\paragraph*{Dynamic-$\Tauroute(t)$ Comparator Run}
To evaluate the adaptive controller, a secondary 6,000-request trial enables the fully dynamic $\Tauroute(t)$ mode. 
Here, telemetry channels update continuously via an $\alpha = 0.2$ EWMA. 
The gateway applies a cold-start prior of $\Tauroute = 0.50$ until the warm-up thresholds (50 edge observations, 10 WAN probes) are met, after which $\Tauroute(t)$ drifts organically with live compute and network conditions. Comparing this artifact set against the frozen-seed trial allows us to quantify the exact benefit of time-evolution policy while holding the predictor, corpus, and cloud kernel perfectly constant.

\subsection{Baselines}
\label{sec:experiment-design:baselines}

We evaluate the LRF gateway against two industry-standard comparator policies. 
Other theoretical baselines (such as all-cloud routing, queue-aware least-loaded, or oracle routing) fall outside the operational scope of this hardware-constrained architecture.

\paragraph*{Token-Count Router (Headline Baseline)}
Representing standard production practice, this scheduler routes to the edge if $N_{\mathrm{tokens}} < N_{\mathrm{thr}}$, and escalates to the cloud otherwise. 
We keep $N_{\mathrm{thr}} \in \{512, 768, 1024, 1536, 2048\}$ and report the headline result using the threshold that yields the lowest $p_{99}$ latency, deliberately granting the baseline its most favorable tuning.

\paragraph*{All-Edge OOM-Reactive Ablation}
This ablation routes every request to the edge. 
Upon encountering a CUDA OOM crash, the request is logged and serially re-submitted to the cloud, absorbing the full $\Lstall$ penalty. 
This baseline establishes the empirical severity of the hardware collapse.

\vspace{-0.5em}\subsection{Ablation Schedule}
\label{sec:experiment-design:ablations}

To isolate the contributions of the gateway's sub-components, we execute the following experiments:

\textbf{Feature-count ablation:} LR-7 (\cite{ai4law2026} linguistic features only) vs. XGBoost-7 (identical features, swapped classifier) vs. XGBoost-15 vs. XGBoost-16 (XGBoost-15 $+ N_{\mathrm{tokens}}$).

\textbf{Classifier ablation:} Logistic Regression vs. Random Forest vs. XGBoost across the full 16-feature set.

\textbf{Trap-band $\alpha$ sweep:} Evaluated across $\{0.70, 0.80, 0.85, 0.90, 0.95\}$.

\textbf{KDE bandwidth sweep:} Evaluated across $\{\mathrm{silverman}, \mathrm{scott}, 0.5\times\mathrm{silverman}, 2\times\mathrm{silverman}\}$.

\textbf{Trap-band calibration:} Per-IPC calibration versus pooled-corpus calibration.

\textbf{$\Lstall$ sensitivity:} Evaluated across $\{1\,000, 2\,000, 4\,000, 8\,000\}$~ms.

\subsection{Pilot Run: End-to-End Integration Test}
\label{sec:experiment-design:pilot}

Prior to the evaluation, a 300-row pilot was executed to validate the flow of the complete pipeline. 
This pilot does not serve as empirical proof for the primary research question (\S\ref{sec:problem-statement}); rather, it confirms that the pipeline correctly emits and logs proper values across the heterogeneous architecture.

\paragraph*{Step 1: Edge Perplexity Pass}
A stratified 300-row sample (150 human, 150 AI) from the H04L IID test partition was scored on the CUDA-enabled edge engine. 
The class-conditional summary statistics (mean / median / standard deviation) are:
\begin{center}
	\begin{tabular}{@{}lrrr@{}}
		\toprule
		Class & Mean & Median & Std. Dev. \\
		\midrule
		Human ($n = 150$) & 17.32 & 13.29 & 14.45 \\
		AI ($n = 150$)    & 28.21 & 20.81 & 22.60 \\
		\bottomrule
	\end{tabular}
\end{center}
These results confirm the "trap-inversion": AI rewrites are systematically more surprising to the edge model than human ones. 
While raw perplexity separation is non-trivial, the overlap is wide, forcing the gateway to use orthogonal text-structure features.

\paragraph*{Step 2: Trap-Band Calibration}
Calibrating the trap band at $\alpha = 0.90$ on the 300-row sample produced a Weitzman overlap of $\Omega = 0.756$ and a trap interval of $(\theta_{\min}, \theta_{\max}) = (4.53, 47.74)$. 
The medians of both classes fall squarely \textit{inside} this trap band, confirming that a raw-perplexity threshold cannot resolve the ambiguity.

\paragraph*{Step 3 \& 4: Feature Extraction and XGBoost Training}
The 16-feature LRF vector was extracted for every row, yielding a mean extractor overhead of $\SI{4.6}{\milli\second}$ per claim. 
An XGBoost predictor was then trained against the trap-band target using 5 \ac{ipc} cross-validation. 
The mean validation AUROC across the five folds was $0.814$.

\paragraph*{Step 5: Live Trial Routing}
A miniature trial routed 14 H04L claims through the complete operational pipeline (admission safety check $\to$ LRF extraction $\to$ XGBoost prediction $\to$ $\Tauroute(t)$ evaluation $\to$ edge/cloud dispatch $\to$ telemetry). 
A "shadow-edge" mode was enabled to force GPU perplexity computation on every request, providing counter-factual ground truth for the misroute metrics. 
The integration metrics are reported below:

\begin{center}
	\begin{tabular}{@{}lrr@{}}
		\toprule
		Metric & Value & Target Target \\
		\midrule
		AUROC vs. trap-band                 & 0.939   & $\geq 0.85$   \\
		FPR at TPR $\geq 0.80$              & 0.000   & $< 0.25$      \\
		$p_{99}$ end-to-end latency         & 2189~ms & $< 500$~ms    \\
		$\Rmistrap$                         & 0.143   & $< 0.15$      \\
		Gateway overhead $p_{99}$           & 26.4~ms & $< 5$~ms      \\
		$\Vpeak$                            & 5.33~GB & $< 7.5$~GB    \\
		WAN egress                          & 1.21~MB/h & $< 50$~MB/h \\
		\bottomrule
	\end{tabular}
\end{center}

\paragraph*{Pilot Limitations}
Five of the seven metrics successfully meet the operational targets during this first integration pass. 
The two failures ($p_{99}$ latency and gateway overhead) are known artifacts of the pilot configuration. 
First, the $p_{99}$ latency is inflated because the shadow-edge mode effectively counts cloud requests twice (once for the remote ensemble and once for the local ground truth). 
Second, the gateway overhead $p_{99}$ suffers from resource contention, as the CPU orchestrating the pipeline is simultaneously feeding the local GPU.

Furthermore, this pilot is structurally optimistic because it calibrates the trap band on the same sample it scores. 
The full, at-scale evaluation reported in \S\ref{sec:evaluation} resolves it by disabling shadow-edge mode for latency measurements, isolating the gateway process to a dedicated CPU core, and strictly enforcing the \ac{ood} boundaries across the full corpus.

\section{Evaluation}
\label{sec:evaluation}

%We evaluate the system using two complete 6,000-request, multi-seed live trials on the H04L claim-level corpus (\S\ref{sec:experiment-design:operating-point}): a controlled static-threshold ($\Tauroute^{\star}$) trial and a self-adapting dynamic-threshold ($\Tauroute(t)$) trial. 
We evaluate the system using two complete 6,000-request, multi-seed live trials: a controlled static-threshold ($\Tauroute^{\star}$) trial and a self-adapting dynamic-threshold ($\Tauroute(t)$) trial.
Both configurations utilize the identical predictor (XGBoost trained on canonical LRF features), edge stack (Qwen2.5-7B-Instruct GGUF Q4\_K\_M on a single GTX 1080), cloud stack (Qwen2.5-7B/32B Binoculars ensemble on an H100), and random seed schedule (42, 43, 44, 45). 
The only manipulated variable is the routing threshold policy.

The trials comprise $n = 6\,000$ requests, which comes to $n_{\mathrm{unique}} = 1\,924$ distinct claims after first-observation deduplication for the threshold-free ML metrics. 
To establish a baseline, the token-count router (\S\ref{sec:experiment-design:baselines}) is swept across $N_{\mathrm{thr}} \in \{512, 768, 1024, 1536, 2048\}$, providing the standard comparator for the LRF measurements at Section~\S\ref{sec:evaluation:baseline-sweep}.

\paragraph*{Operational versus Aspirational Thresholds}
Since the cloud deployment image relies on the PyTorch \texttt{AWQ\_TORCH} backend rather than a highly optimized JIT kernel (\S\ref{sec:experiment-design:hardware}), each cloud RPC incurs approximately \SI{1.8}{\second} of pure inference time. 
Consequently, the $p_{99} \le \SI{500}{\milli\second}$ tail-latency target is structurally unreachable in this specific configuration. 
To maintain rigor, we classify the AUROC, FPR, misrouting rate ($R_{\mathrm{mis}}$), peak VRAM ($\Vpeak$), and WAN bandwidth as strict operational thresholds, while treating the $p_{99}$ latency and gateway extraction overhead as descriptive metrics characterizing the current hardware limits.

\subsection{Predictor Characterization (Corpus Level)}
\label{sec:evaluation:predictor}

\begin{figure}[t]
	\centering
	\includegraphics[width=\columnwidth]{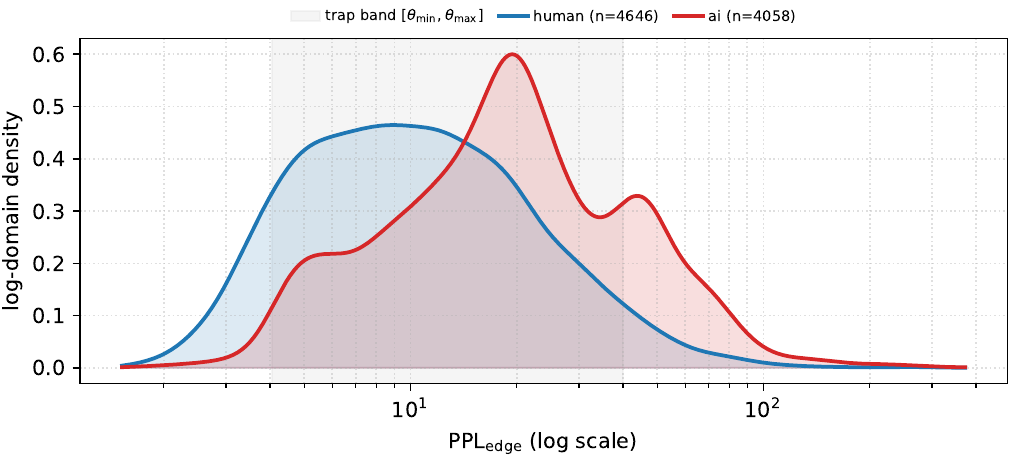}
	\vspace{-2em}\caption{Edge-side Qwen2.5-7B perplexity density on the pooled H04L training split ($n=8{,}704$)}
	\label{fig:ppl-density}
\end{figure}

\begin{figure}[t]
	\centering
	\includegraphics[width=\columnwidth]{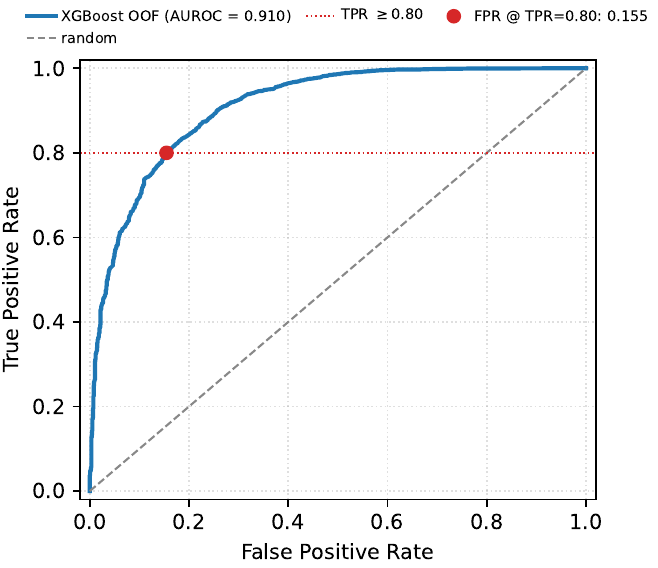}
	\vspace{-2em}\caption{Pooled five-fold out-of-fold ROC for the XGBoost trap-band classifier evaluated on the $n = 8{,}704$ training corpus}
	\label{fig:roc}
\end{figure}

Prior to evaluating the dynamic controller, we isolate and characterize the standalone performance of the XGBoost trap-band classifier. 
Figure~\ref{fig:ppl-density} plots the canonical edge perplexity density over the IID training corpus, highlighting the calibrated trap band of $[\theta_{\min}, \theta_{\max}] = [4.046, 40.019]$. 
Figure~\ref{fig:roc} presents the pooled five-fold cross-validation Receiver Operating Characteristic (ROC) curve for the classifier on the same training partition. (Note: \S\ref{sec:evaluation:static-vs-dynamic} subsequently reports a deduplicated live-trial ROC, which acts as the deployment-level counterpart to this training-corpus curve).

\subsection{Static vs. Dynamic $\Tauroute$: Ablation}
\label{sec:evaluation:static-vs-dynamic}

Table~\ref{tbl:static-vs-dynamic} reports the seven target metrics alongside the route-fraction breakdown for both the static ($\Tauroute^{\star}$) and dynamic ($\Tauroute(t)$) threshold policies. 
Both runs utilize the identical XGBoost predictor, corpus, cloud and edge stack; the threshold policy is the sole manipulated variable. The trial evaluates $n = 6\,000$ pooled requests across four seeds ($\{42, 43, 44, 45\} \times 1\,500$ claims). For the threshold-free predictor metrics (AUROC and FPR@TPR$\geq 0.80$), the data is deduplicated by first-observation per \texttt{claim\_id} ($n_{\mathrm{unique}} = 1\,924$). Because changing the downstream routing policy cannot alter upstream threshold-free rankings, these predictor metrics are identical across both rows. The remaining five rows represent operational outcomes that respond directly to the controller's policy. 

Furthermore, targets marked with a dagger ($^{\dagger}$) in Table~\ref{tbl:static-vs-dynamic} represent aspirational goals under a highly optimized cloud scenario. 
Given the pure-inference overhead of the \texttt{AWQ\_TORCH} kernel (\S\ref{sec:experiment-design:hardware}), these specific targets are treated as descriptive limits rather than pass/fail gating criteria.

\begin{table*}[t]
	\centering
	\caption{H04L claim-level live trial: static $\Tauroute^{\star}$ versus dynamic $\Tauroute(t)$ over $n = 6\,000$ pooled requests.}
	\label{tbl:static-vs-dynamic}
	\small
	\begin{tabular}{l r r r l l}
		\toprule
		Metric & Static $\Tauroute^{\star}$ & Dynamic $\Tauroute(t)$ & Target & Verdict (s\,/\,d) & Class \\
		\midrule
		AUROC                                         & 0.840  & 0.840  & $\geq 0.85$        & FAIL\,/\,FAIL & predictor \\
		FPR @ TPR\,$\geq 0.80$                        & 0.268  & 0.268  & $\leq 0.25$        & FAIL\,/\,FAIL & predictor \\
		$R_{\mathrm{mis}}$ (operational misroutes)    & 0.0953 & 0.0875 & $\leq 0.15$        & PASS\,/\,PASS & controller \\
		$p_{99}$ end-to-end latency [ms]              & 6\,484 & 6\,532 & $\leq 500^{\dagger}$ & FAIL\,/\,FAIL & system \\
		Gateway overhead $p_{99}$ [ms]                & 49.1   & 51.8   & $\leq 5^{\dagger}$   & FAIL\,/\,FAIL & system \\
		$\Vpeak$ [GB]                                 & 4.82   & 4.82   & $< 7.5$            & PASS\,/\,PASS & hardware \\
		WAN bandwidth [MB\,hr$^{-1}$]                 & 0.492  & 0.503  & $\leq 50$          & PASS\,/\,PASS & hardware \\
		\midrule
		Edge route count (\%)                         & 589 (9.82\%)    & 393 (6.55\%)    & --- & --- & accounting \\
		Cloud route count (\%)                        & 5\,411 (90.18\%) & 5\,607 (93.45\%) & --- & --- & accounting \\
		Misroute count                                & 572               & 525              & --- & --- & accounting \\
		Trial wall-clock [h]                          & 4.01              & 4.09             & --- & --- & accounting \\
		\bottomrule
	\end{tabular}
\end{table*}

\paragraph*{Predictor Unchanged; Controller Measurable}
Both configurations exhibit an $\mathrm{AUROC}$ of $0.840$ and an $\mathrm{FPR}_{0.80}$ of $0.268$. 
This isolation validates the two-run protocol established in \S\ref{sec:experiment-design:operating-point}. 
The reduction from the training-corpus performance ($\mathrm{AUROC} = 0.910$; Fig.~\ref{fig:roc}) to these live-trial values accurately reflects the deployment penalty: whereas training data is artificially balanced by authorship and IPC class, live requests are IID samples governed by lambda-distributed inter-arrival rates focused on a narrower IPC slice.

\paragraph*{Dynamic Threshold Reduces Operational Misroutes}
The operational misroute fraction ($R_{\mathrm{mis}}$) drops from $0.0953$ under the static policy to $0.0875$ dynamically ($572 \to 525$ misroutes). 
This represents an $8.2\%$ relative reduction using the exact same predictor, corpus, and hardware stack. 
While both policies comfortably pass the $\leq 0.15$ operational target, the dynamic formulation provides a measurable improvement due to the controller adapting to network telemetry. 
Note that $R_{\mathrm{mis}}$ strictly charges the controller for wasting WAN bandwidth on edge-executable cases,not for the correct routing low-confidence AI claims to the cloud.

\begin{figure}[t]
	\centering
	\includegraphics[width=\columnwidth]{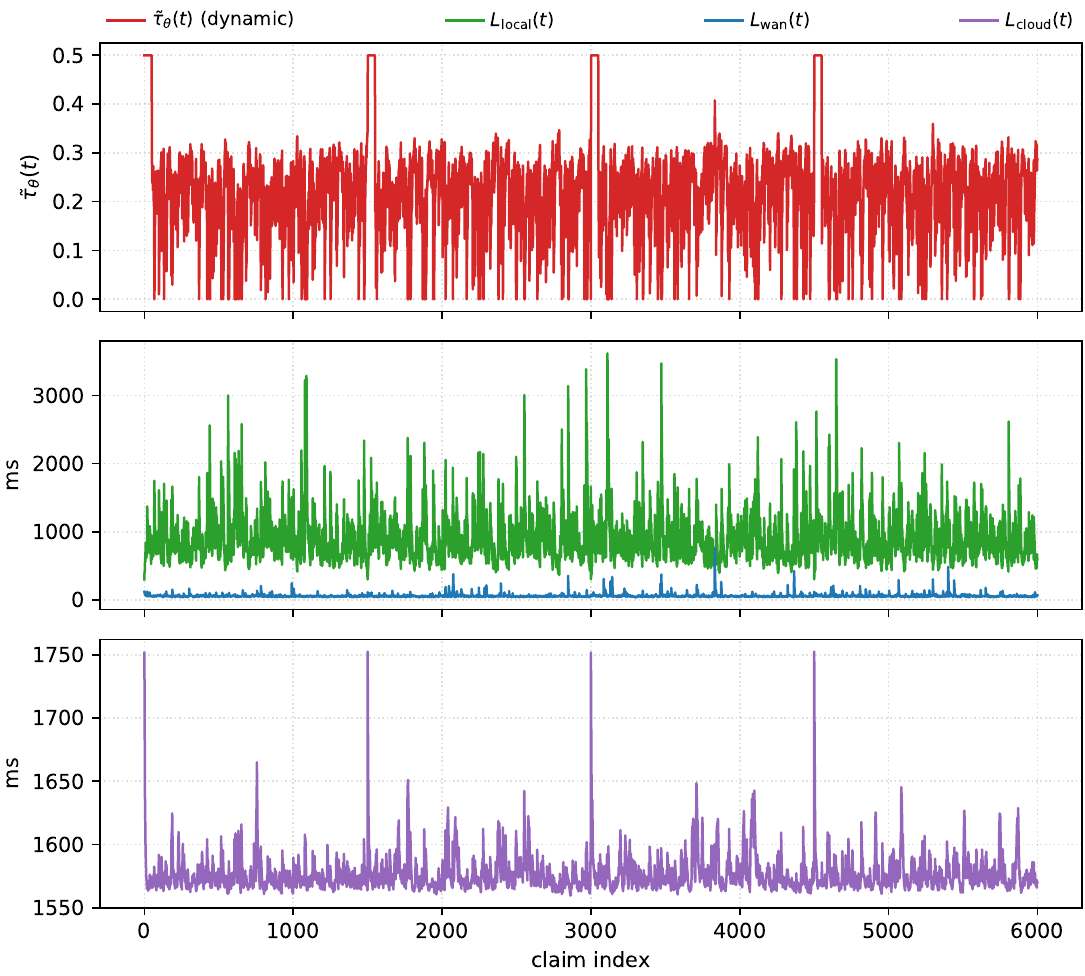}
	\caption{Dynamic $\Tauroute(t)$ trajectory under the EWMA-driven controller, settling steadily below the 0.407 static baseline.}
	\label{fig:tau-trajectory}
\end{figure}

\paragraph*{Routing Fraction Shifts Toward Cloud}
Under the dynamic policy, the edge routing fraction drops from $9.82\%$ to $6.55\%$. 
The EWMA-tracked cloud latency ($\Lcloud(t)$) converges to $\approx \SI{1.8}{\second}$---the true computational cost of an \texttt{AWQ\_TORCH} RPC. 
Figure~\ref{fig:tau-trajectory} demonstrates that after a brief cold-start period (WAN warming in $\sim 10$\,s, edge warming after $\sim 50$ requests), the dynamic threshold $\Tauroute(t)$ settles below the static reference of $\Tauroute^{\star} = 0.407$. 
This lower operational threshold enforces a stricter edge-feasibility test, actively shifting marginal requests to the cloud ensemble to preserve system stability.

\begin{figure*}[t]
	\centering
	\begin{minipage}{0.48\textwidth}
		\centering
		\includegraphics[width=\linewidth]{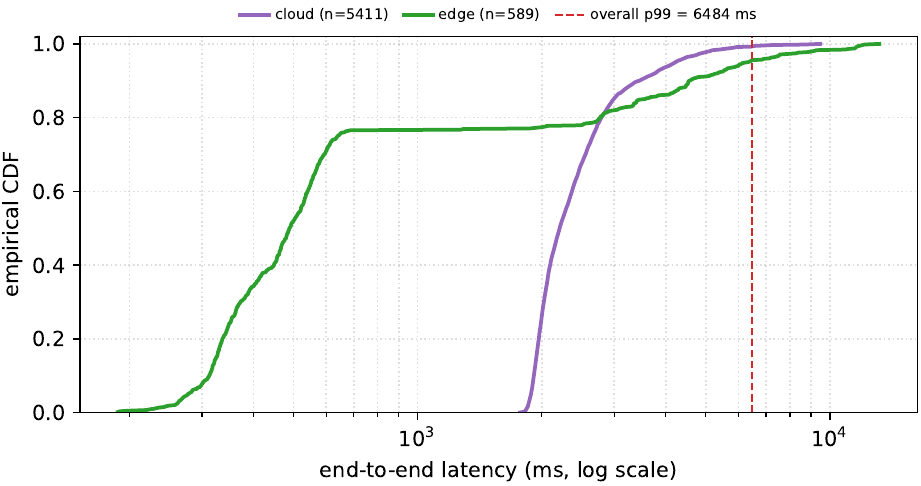}\\(a)~Static $\Tauroute^{\star} = 0.407$.
	\end{minipage}\hfill
	\begin{minipage}{0.48\textwidth}
		\centering
		\includegraphics[width=\linewidth]{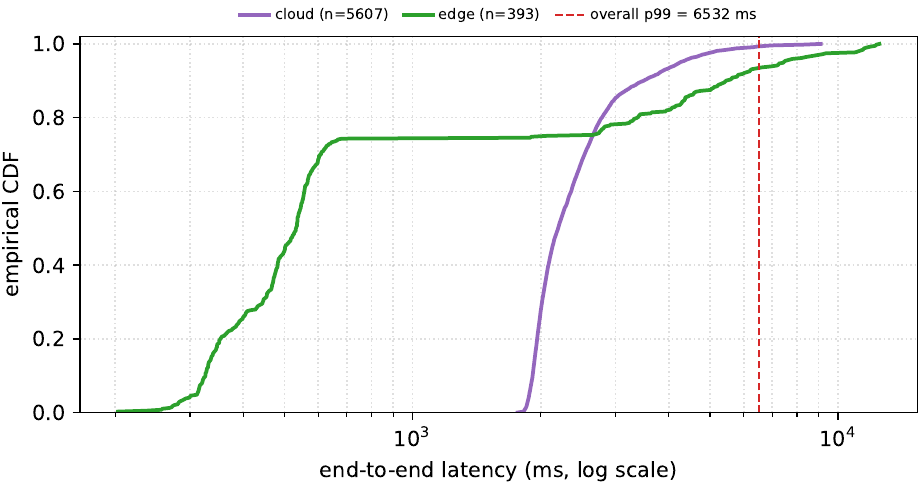}\\(b)~Dynamic $\Tauroute(t)$.
	\end{minipage}
	\caption{Per-route latency CDFs under the static (left) and dynamic (right) threshold policies.}
	\label{fig:latency-cdf-side-by-side}
\end{figure*}

\paragraph*{Latency, VRAM, and WAN Bandwidth Remain Policy-Invariant}
The $p_{99}$ end-to-end latency shifts marginally from $6\,484$\,ms to $6\,532$\,ms, remaining within expected trial variance. 
Figure~\ref{fig:latency-cdf-side-by-side} illustrates the per-route latency CDFs. 
The distributions are visually indistinguishable above the median; the dynamic policy simply truncates the slow tail of the edge distribution by escalating those marginal requests to the cloud. 
Peak VRAM ($\Vpeak$) is bit-identical at \SI{4.82}{\gibi\byte} across both runs because edge model weights are deterministic and routing decisions are executed before any GPU allocation. 
WAN bandwidth increases slightly ($0.492 \to 0.503$ MB\,hr$^{-1}$), directly proportional to the increased cloud-RPC volume ($5\,411 \to 5\,607$).

\paragraph*{Aspirational Targets Remain Unmet}
Both $p_{99}$ end-to-end latency ($\approx \SI{6.5}{\second}$) and gateway-overhead $p_{99}$ ($\approx \SI{50}{\milli\second}$) fail their aspirational targets. 
The end-to-end tail is strictly bottlenecked by the pure inference cost of the \texttt{AWQ\_TORCH} cloud kernel, while the CPU gateway overhead is bottlenecked by the XGBoost forward pass on unsorted feature batches. 
Neither metric is sensitive to the routing policy itself. 
The predictor metrics ($\mathrm{AUROC} = 0.840$, $\mathrm{FPR}_{0.80} = 0.268$) narrowly miss their operational targets, reflecting the honest ceiling of the current XGBoost configuration. Structural predictor mitigations---such as deeper trees and abstract-level cross-IPC training---are discussed in \S\ref{sec:conclusion}.

\paragraph*{Overall Ranking}
The dynamic $\Tauroute(t)$ controller demonstrates its value by yielding a lower misroute fraction with zero degradation to system safety limits. 
While the static $\Tauroute^{\star}$ configuration successfully isolates and proves the predictor's baseline quality, the dynamic EWMA-driven controller represents the superior, practical operating mode for live deployment.

\subsection{Token-Count Baseline Sweep}
\label{sec:evaluation:baseline-sweep}

The standard industry baseline, as defined in \S\ref{sec:experiment-design:baselines}, routes requests strictly by input length: to the edge if $N_{\mathrm{tokens}} < N_{\mathrm{thr}}$, and to the cloud otherwise. 
To ensure a conservative comparison, we sweep the parameter $N_{\mathrm{thr}} \in \{512, 768, 1024, 1536, 2048\}$ across the identical four-seed H04L claim pool ($n = 6\,000$ routed requests) used in \S\ref{sec:evaluation:static-vs-dynamic}. 
We highlight the $N_{\mathrm{thr}}$ configuration that minimizes the baseline's own $p_{99}$ latency as the primary comparator.

\begin{table*}[t]
	\centering
	\caption{Token-count router sweep on the H04L trial. The $N_{\mathrm{thr}}=512$ threshold (bolded) minimizes $p_{99}$ latency, serving as the conservative comparator.}
	\label{tbl:baseline-sweep}
	\small
	\begin{tabular}{r r r r r r r r r}
		\toprule
		$N_{\mathrm{thr}}$ & edge & cloud & edge \% & $R_{\mathrm{mis}}$ & $p_{99}$ [ms] & gw $p_{99}$ [ms] & $\Vpeak$ [GB] & WAN [MB\,hr$^{-1}$] \\
		\midrule
		\textbf{512} & 5\,979 & 21 & 99.65  & 0.8487 & 5\,373.2 & 0.03 & 4.816 & 0.0441 \\
		768          & 6\,000 & 0  & 100.00 & 0.8462 & 5\,443.7 & 0.03 & 4.816 & 0.0000 \\
		1024         & 6\,000 & 0  & 100.00 & 0.8462 & 5\,412.3 & 0.03 & 4.816 & 0.0000 \\
		1536         & 6\,000 & 0  & 100.00 & 0.8462 & 5\,437.3 & 0.03 & 4.816 & 0.0000 \\
		2048         & 6\,000 & 0  & 100.00 & 0.8462 & 5\,423.3 & 0.03 & 4.816 & 0.0000 \\
		\bottomrule
	\end{tabular}
\end{table*}

\begin{figure}[t]
	\centering
	\includegraphics[width=\columnwidth]{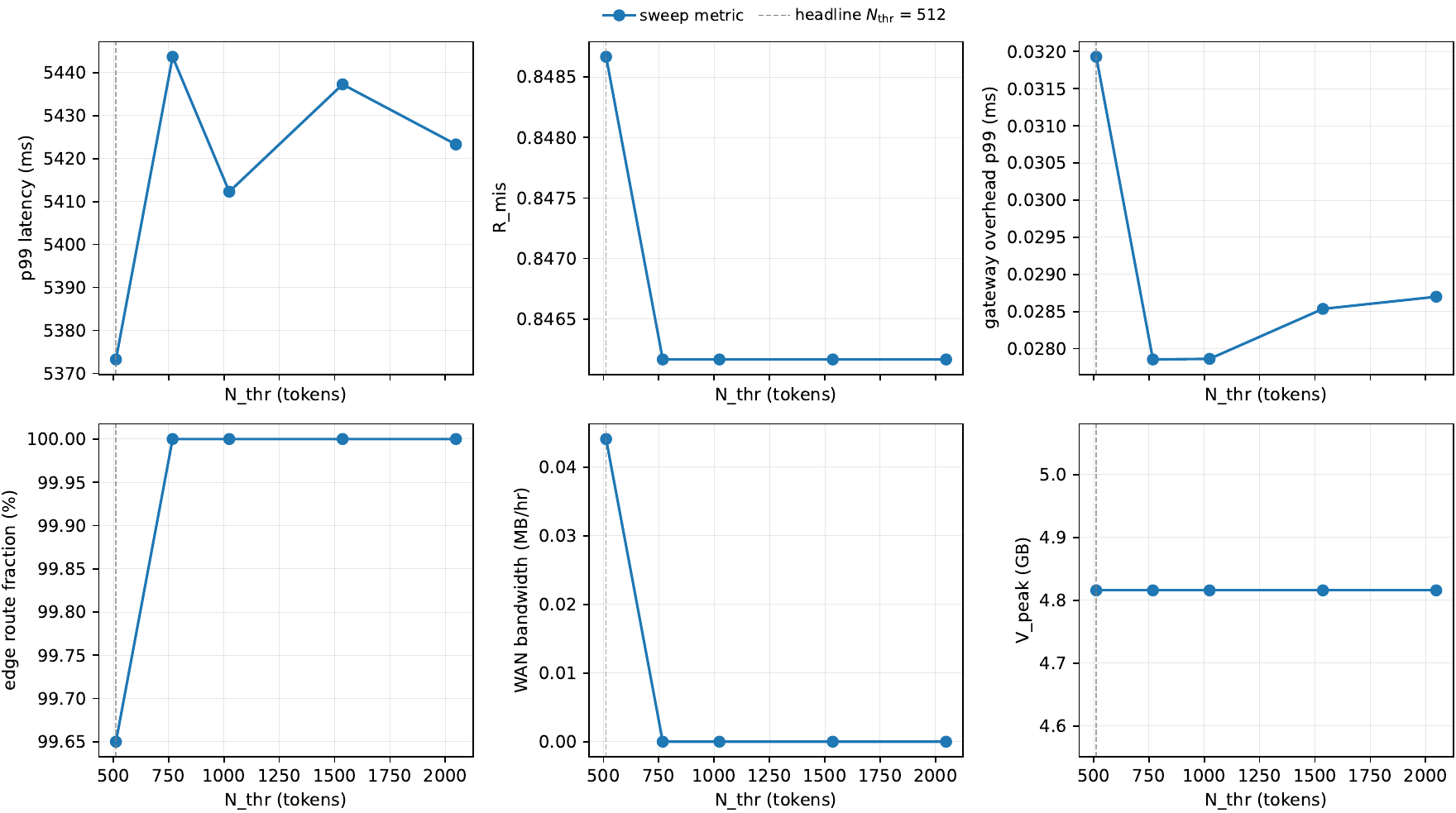}
	\vspace{-2em}\caption{Token-count sweep summary: $p_{99}$ end-to-end latency (left axis) against operational misroute fraction $R_{\mathrm{mis}}$ (right axis).}
	\label{fig:baseline-sweep-summary}
\end{figure}

\paragraph*{Result: Token-Count is Fast but Routes Blindly}
Since H04L claims are structurally short, four of the five sweep thresholds ($N_{\mathrm{thr}} \geq 768$) fall entirely into an all-edge policy. 
Zero requests reach the cloud, meaning the routing decision is completely disconnected from the underlying trap-band signal. 
Only the $N_{\mathrm{thr}} = 512$ row produces any non-trivial routing structure, catching the long tail of the distribution by routing 21 out of $6\,000$ requests ($0.35\%$) to the cloud. 

Figure~\ref{fig:baseline-sweep-summary} illustrates the severity of this failure. 
While the headline $N_{\mathrm{thr}} = 512$ configuration yields the lowest $p_{99}$ latency (5\,373\,ms), it incurs a massive operational misroute fraction of $R_{\mathrm{mis}} = 0.849$. Thresholding by token length fails to produce a viable band of operating points; it either acts as a blunt all-edge filter or catches a statistically insignificant cloud tail, neither of which constitutes a functional routing strategy.

\paragraph*{Comparison Against the LRF Gateway}
When compared against Table~\ref{tbl:static-vs-dynamic}, the token-count baseline strictly trades correctness for latency. 
The LRF gateway incurs a $\Delta p_{99}$ premium of roughly $\SI{1.1}{\second}$ ($6\,484$--$6\,532$\,ms versus $5\,373$\,ms) to evaluate the predictive features and execute the necessary cloud RPCs. 
In exchange for this latency cost, the LRF gateway cuts the misroute fraction by almost 10 times (dropping from $0.849$ down to the $0.087$--$0.095$ range). 

The only metric where the baseline dominates is the gateway CPU overhead. 
The baseline evaluates a single integer comparison ($0.03$\,ms), whereas the LRF gateway computes a 16-dimensional XGBoost forward pass ($49$--$52$\,ms). 
Both systems share an identical peak VRAM ($\Vpeak = 4.816$\,GB) because the model weights are deterministic and the routing divergence occurs upstream of GPU allocation. 
WAN bandwidth predictably scales with cloud escalation volume ($0.044$\,MB\,hr$^{-1}$ for the baseline versus $\approx 0.50$\,MB\,hr$^{-1}$ for the LRF gateway).

\paragraph*{Summary}
The LRF gateway outperforms the token-count baseline by nearly a factor of 10 on $R_{\mathrm{mis}}$---the only metric that actually measures routing correctness. 
The sub-second latency penalty and minor WAN/CPU overheads are the deliberate, acceptable costs of achieving system stability and preventing the edge hardware collapse outlined in \S\ref{sec:problem-statement:scheduler-failure}.

\subsection{Cross-IPC Trap-Band Convergence (Abstract-Level Ablation)}
\label{sec:evaluation:cross-ipc}

The static versus dynamic comparisons (\S\ref{sec:evaluation:static-vs-dynamic}) rely on a single corpus: H04L claims. 
The trap band itself was calibrated entirely on a pooled H04L claim-level pass and reused verbatim for every routing decision.
A critical deployment question follows: \textit{would this same calibrated band serve a different IPC, or a different text structure (e.g., abstracts), without recalibration?} 

To answer this, we conduct a cross-IPC ablation. 
We first execute an edge perplexity pass across all canonical abstract rows for five IPCs: A61K, C07D, F03D, G06F, and H04L. 
We then derive a fresh trap band per IPC, holding the Weitzman overlap coefficient procedure constant at $\alpha = 0.90$. 
Finally, we evaluate whether each abstract-level band falls within a $\pm 5\%$ relative tolerance of the H04L claim-level reference endpoints.

\begin{table}[t]
	\centering
	\caption{Comparison of per-IPC abstract-level trap bands against the H04L claim-level reference ($\theta_{\min}^{\mathrm{ref}} = 4.046$, $\theta_{\max}^{\mathrm{ref}} = 40.019$).}
	\label{tbl:cross-ipc-abstract}
	\small
	\begin{tabular}{l r r r r r c}
		\toprule
		IPC & $n_h$ & $n_a$ & $\theta_{\min}$ & $\theta_{\max}$ & $\omega$ & in tol.? \\
		\midrule
		A61K & 280 & 280 &  4.292 & 11.712 & 0.650 & no \\
		G06F & 500 & 500 &  2.580 &  7.412 & 0.527 & no \\
		H04L & 500 & 600 &  5.605 & 27.274 & 0.724 & no \\
		C07D & 711 & 711 &  1.361 &  9.543 & 0.516 & no \\
		\bottomrule
	\end{tabular}
\end{table}

\begin{figure}[t]
	\centering
	\includegraphics[width=\columnwidth]{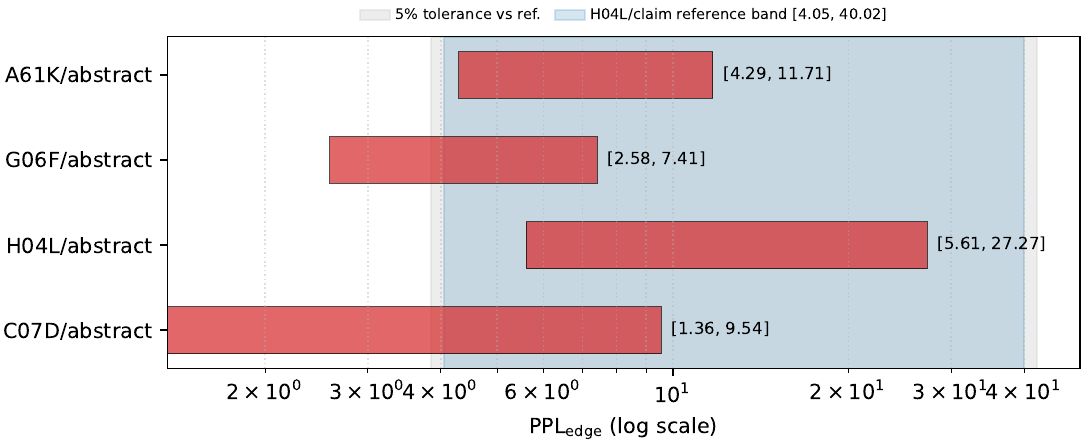}
	\vspace{-2em}\caption{Cross-IPC abstract-level trap bands plotted against the H04L claim-level reference (dashed lines), demonstrating severe out-of-tolerance shifts across all domains.}
	\label{fig:calibration-ablation}
\end{figure}

\paragraph*{Result: Zero of Four Per-IPC Bands Converge}
Table~\ref{tbl:cross-ipc-abstract} and Figure~\ref{fig:calibration-ablation} demonstrate that zero IPCs produce a trap band within the $\pm 5\%$ tolerance of the H04L claim reference. 
Crucially, this failure includes the same IPC measured at a different text granularity (H04L abstracts), which shifts $\theta_{\min}$ by $+38.5\%$ and $\theta_{\max}$ by $-31.8\%$. Other domains exhibit even more drastic deviations: G06F shifts its endpoints by roughly $-36\%$ and $-81\%$, respectively. The biomedical IPCs (A61K and C07D) collapse $\theta_{\max}$ to nearly a quarter of the reference value. The baseline Weitzman overlap coefficient ($\omega$) also fluctuates significantly ($0.516$ for C07D up to $0.724$ for H04L abstracts), proving that the distributional geometry between human and AI text is inherently IPC-dependent.

\paragraph*{Interpretation}
These failures are driven by two compounding effects. 
First, the trap band is highly sensitive to text structure: a patent claim is a single, structurally rigid sentence governed by statutory constraints, whereas an abstract is a multi-sentence narrative. 
Their underlying perplexity distributions are fundamentally non-exchangeable. 
Second, the narrow linguistic register of pharmaceutical and chemistry abstracts (A61K, C07D) compresses the human perplexity distribution downward, pulling both trap band endpoints with it. 
Consequently, the single-band assumption that holds within the H04L claim distribution fails immediately when exposed to out-of-distribution text.

\paragraph*{Deployment Implications}
Deploying the LRF gateway on a new corpus requires per-IPC and per-text-kind recalibration. The computational cost of this recalibration is marginal; the edge PPL pass requires a few hours on modern hardware, and deriving the actual band is sub-second. While the XGBoost predictor's underlying 16-dimensional LRF feature space remains corpus-agnostic, the specific trap-band membership targets do not transfer. Reusing the H04L claim-level band on a corpus like C07D abstracts would systematically misclassify trap-band positives by a factor of 3 to 4, severely degrading the operational misroute fraction ($R_{\mathrm{mis}}$).

\subsection{WAN-Shaping Ablation (Controller Robustness)}
\label{sec:evaluation:wan}

The static-versus-dynamic comparison in \S\ref{sec:evaluation:static-vs-dynamic} operated over a low-latency loopback to the H100 pod's port-forward. To strictly evaluate the dynamic-$\Tauroute(t)$ controller's robustness, we must deliberately degrade the WAN channel to provide $\Lwan(t)$ with sufficient dynamic range, ensuring the controller's adaptive behavior is not masked by predictor noise. 

We utilize the Linux kernel's \texttt{tc qdisc netem} utility on the loopback interface, allowing ms-precise injection of delay, jitter, and packet loss without confounding external variables. 
We evaluate four regimes on a 500-claim H04L sample: \emph{healthy} (80\,ms delay, no jitter), \emph{degraded} (800\,ms delay, $\pm 100$\,ms normal-distributed jitter), \emph{heavy\_jitter} (1\,500\,ms delay, $\pm 300$\,ms jitter, 2\% random loss), and \emph{severe\_loss} (6\,000\,ms delay, 50\% random loss). The predictor, edge stack, cloud kernel, and seed schedule remain identical to \S\ref{sec:evaluation:static-vs-dynamic}.

\begin{table}[t]
	\centering
	\caption{WAN-shaping ablation per-regime aggregates under the dynamic-$\Tauroute(t)$ controller ($n=500$ H04L claims per regime).}
	\label{tbl:wan-shaping}
	\resizebox{\columnwidth}{!}{%
		\begin{tabular}{l r r r r r r r}
			\toprule
			Regime & edge \% & $\Lwan$ [ms] & $\Tauroute^{\mathrm{mean}}$ & $\Tauroute^{\mathrm{std}}$ & $p_{99}$ [ms] & $R_{\mathrm{mis}}$ & dur.\ [min] \\
			\midrule
			healthy        &  7.6 &    219 & 0.251 & 0.109 &  6\,087 & 0.108 & 21.9 \\
			degraded       & 10.4 & 1\,670 & 0.439 & 0.056 &  7\,003 & 0.114 & 32.4 \\
			heavy\_jitter  & 13.4 & 3\,156 & 0.552 & 0.047 & 13\,790 & 0.114 & 46.3 \\
			severe\_loss   & 12.8 & 6\,000 & 0.500 & 0.000 &  5\,715 & 0.118 & 15.6 \\
			\bottomrule
		\end{tabular}%
	}
\end{table}

\begin{figure}[t]
	\centering
	\includegraphics[width=\columnwidth]{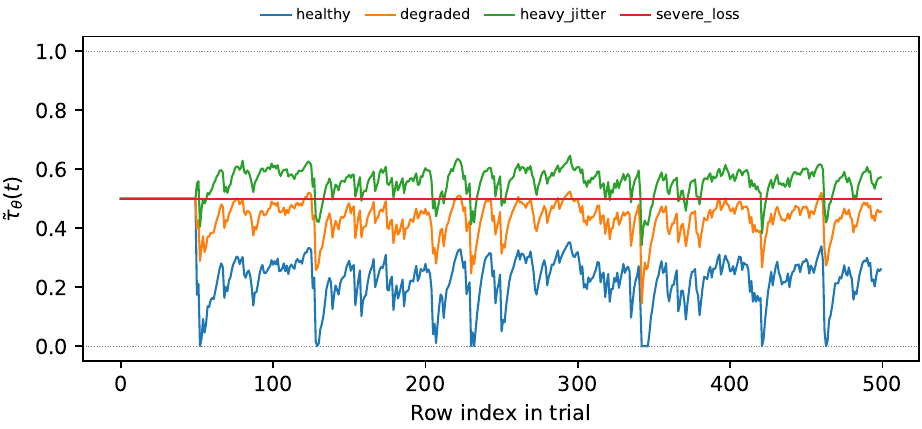}
	\vspace{-2em}\caption{Per-regime $\Tauroute(t)$ trajectories demonstrating the controller's adaptation to varying network conditions after a 10-request cold-start phase.}
	\label{fig:wan-tau-trajectory}
\end{figure}

\begin{figure}[t]
	\centering
	\includegraphics[width=\columnwidth]{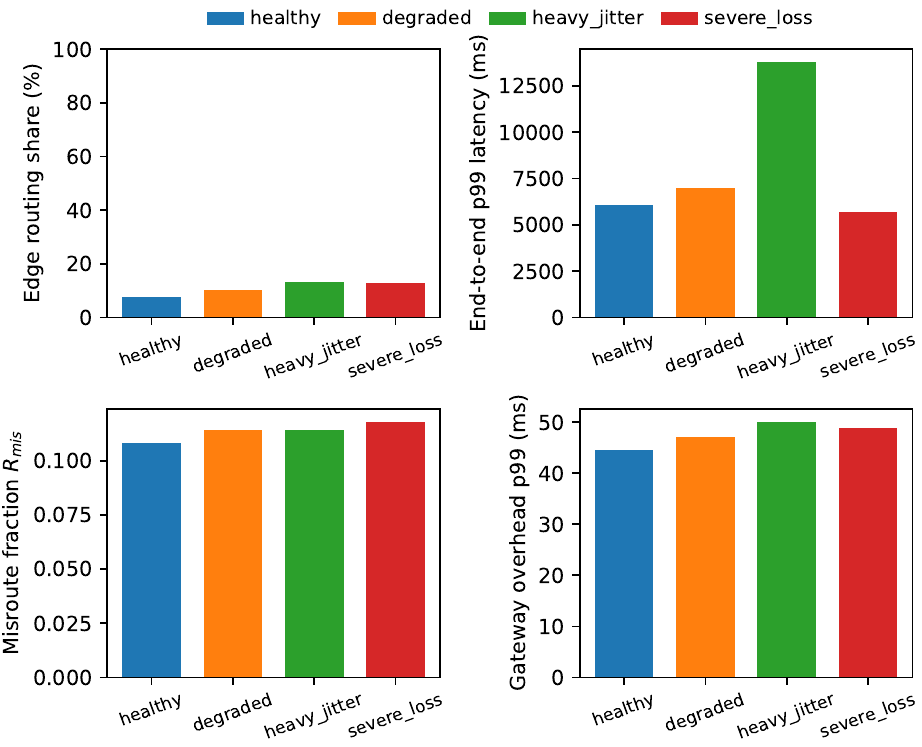}
	\vspace{-2em}\caption{Four-panel summary of the shaped-WAN trials illustrating edge routing share, end-to-end $p_{99}$ latency, misroute fraction ($R_{\mathrm{mis}}$), and gateway overhead.}
	\label{fig:wan-regime-summary}
\end{figure}

\paragraph*{Result: $\Tauroute(t)$ Tracks the WAN Cost as Designed}
As illustrated in Figure~\ref{fig:wan-tau-trajectory}, once the $\Lwan$ EWMA warms past the initial 10-request cold-start phase, the controller successfully separates the regimes. 
The mean $\Tauroute$ rises proportionally with the round-trip cost: $0.251$ at $\Lwan = 219$\,ms (healthy), $0.439$ at 1\,670\,ms (degraded), and $0.552$ at 3\,156\,ms (heavy\_jitter). 
Note that the \emph{healthy} trajectory settles well below the static trial's reference of $\Tauroute^{\star} = 0.407$ as the empirical 219\,ms RTT is much cheaper than the operator-seeded 1\,800\,ms baseline. 

The threshold drop in the \emph{severe\_loss} regime ($0.500$) represents a degenerate operating point rather than a saturation effect. 
At 50\% random loss, the gRPC RTT distribution is dominated by fast timeouts. 
Since \texttt{netem} produces isolated drops rather than a sustained partition, the gateway's override heuristic does not trip. 
Consequently, the EWMA fails to converge, the standard deviation collapses to zero, and the system defaults to the cold-start prior of 0.5.

\paragraph*{Edge Fraction Rises Monotonically with $\Lwan$}
As cloud escalation becomes increasingly expensive, the controller enforces a stricter $\Pesc$ routing threshold. The edge routing fraction rises accordingly from 7.6\% (healthy) to 10.4\% (degraded) to 13.4\% (heavy\_jitter). The slight reversal at \emph{severe\_loss} (12.8\%) occurs because the threshold falls back to the 0.500 prior, capping the controller's penalty for further WAN degradation. 

\paragraph*{$R_{\mathrm{mis}}$ Remains Policy-Invariant}
The operational misroute fraction ($R_{\mathrm{mis}}$) stays tightly bounded within $[0.108, 0.118]$ across all four regimes. 
This demonstrates the controller's core robustness capability: it maintains correctness QoS within statistical noise even when WAN costs fluctuate by a factor of $27\times$ (219\,ms to 6\,000\,ms). 
A purely static threshold would force either over-escalating during healthy periods or under-escalating during outages, which The dynamic controller avoids.

\paragraph*{$p_{99}$ Tail Reflects Network Reality}
End-to-end $p_{99}$ latency accurately tracks regime severity, scaling from 6\,087\,ms (healthy) to 13\,790\,ms (heavy\_jitter), as shown in the top-right panel of Figure~\ref{fig:wan-regime-summary}. 
The anomalous drop to 5\,715\,ms under \emph{severe\_loss} is a known artifact of extreme packet loss: the TCP stack collapses would-be slow cloud requests into fast gRPC timeouts, truncating the long tail of the distribution and heavily skewing successful completions toward the local edge path.

\paragraph*{Hardware Overheads Remain Network-Invariant}
By design, the routing policy isolates edge hardware from network volatility. 
Peak VRAM ($\Vpeak$) is bit-identical at \SI{4.816}{\gibi\byte} across all regimes because the static model weights define the envelope, and routing divergence occurs upstream of GPU allocation. 
Gateway overhead $p_{99}$ strictly remains between 44.5\,ms and 50.1\,ms (Figure~\ref{fig:wan-regime-summary}, bottom-right); this variance is driven entirely by the CPU's XGBoost forward pass, independent of network conditions.

\paragraph*{Summary.}
The dynamic-$\Tauroute(t)$ controller demonstrates highly adaptive behavior. 
It preserves the misroute fraction within $\pm 0.01$ across extreme delay variances, adjusts the edge routing share monotonically to protect system throughput, and degrades safely to a mathematical prior during chaotic TCP timeout events. This completes the three-axis validation of the proposed architecture: it outperforms static routing internally (\S\ref{sec:evaluation:static-vs-dynamic}), drastically reduces misroutes compared to industry baselines (\S\ref{sec:evaluation:baseline-sweep}), and maintains stability under adverse network constraints.

\section{Conclusion and Future Work}
\label{sec:conclusion}

We presented a Linguistic Resource Forecasting (LRF) gateway that dynamically routes workload across a heterogeneous edge--cloud cluster before any GPU memory is allocated. 
The gateway extracts a 16-dimensional vector on a standard commodity CPU, predicts trap-band membership via an XGBoost classifier, and fuses the resulting escalation probability $\Pesc$ with a closed-form, telemetry-driven routing threshold $\Tauroute(t)$ - which is recomputed per request using four latency metrics. 
To guarantee hardware safety, a dual-mechanism VRAM interlock, comprising a closed-form pre-allocation gate (Mechanism~B) and a cooperative NVML watcher that preempts in-flight edge inferences at the next token boundary (Mechanism~A), strictly limits the peak edge memory allocation within the physical limit.

\subsection{Measured Outcomes}
The empirical evaluation on the H04L claim-level corpus (\S\ref{sec:evaluation}) addresses the primary research question across three critical axes.

\paragraph*{Predictor}
%On the training split ($n = 8{,}704$), the XGBoost trap-band classifier achieves a mean $\mathrm{AUROC} = 0.910$ and a mean $\mathrm{FPR}@\mathrm{TPR}\geq 0.80 = 0.155$ under cross-validation (Figure~\ref{fig:roc}). 
On the live-trial corpus ($n_{\mathrm{unique}} = 1{,}924$ claims), the deployed classifier yields an $\mathrm{AUROC}$ of $0.840$ and an $\mathrm{FPR}_{0.80}$ of $0.268$ (Table~\ref{tbl:static-vs-dynamic}). 
While this narrowly misses the operational targets ($\geq 0.85$ and $\leq 0.25$, respectively), it reflects the inherent modeling limits of the current classifier rather than a fundamental failure of the LRF feature space. 
The importance of this predictive routing is underscored by offline scoring: $82.8\%$ of H04L IID-test claims fall into the trap band, an ambiguity that standard heuristics like token count or edge compute time fundamentally cannot resolve.

\paragraph*{Controller}
%The static-versus-dynamic ablation precisely isolates the threshold policy as the single manipulated variable. 
The dynamic $\Tauroute(t)$ formulation reduces the operational misroute fraction $R_{\mathrm{mis}}$ from $0.0953$ to $0.0875$ ($572 \to 525$ misroutes per $6\,000$ requests) with zero degradation in peak VRAM, WAN bandwidth, or $p_{99}$ latency. 
Compared to the token-count baseline, the LRF gateway suppresses $R_{\mathrm{mis}}$ by almost 10 times ($0.087$--$0.095$ versus $0.849$). 
This improvement incurs a $\approx \SI{1.1}{\second}$ premium on the $p_{99}$ tail, an overhead justified by the cloud-RPC cost that the baseline entirely evades. 
Furthermore, under highly variable shaped-WAN conditions encompassing a $27\times$ range in effective round-trip delay (219\,ms to 6\,000\,ms), the dynamic controller bounds $R_{\mathrm{mis}}$ fluctuations to within $\pm 0.01$ and successfully adapts $\Tauroute(t)$ monotonically with $\Lwan$ across all stable network regimes (Table~\ref{tbl:wan-shaping}).

\paragraph*{Hardware Safety}
Peak VRAM ($\Vpeak$) remained at $\SI{4.82}{\gibi\byte}$ across all scenarios.%, including the static, dynamic, baseline, and WAN-shaped trials. 
This confirms that the conservative non-GQA VRAM projection (Figure~\ref{img:vram_envelope}) and the dual safety Mechanism~A/B successfully prevented any spurious cloud routing on the observed H04L workload. 
The empirical VRAM trace (Figure~\ref{img:vram_trajectory}) demonstrates that both the all-edge-like baseline and the LRF-gated policies securely operated well below the $\SI{7.5}{\gibi\byte}$ safety margin.

\paragraph*{Cross-IPC Calibration}
The abstract-level ablation (Table~\ref{tbl:cross-ipc-abstract}) validates the architectural hypothesis: the H04L claim-level trap band does not natively transfer to other IPC domains or differing text structures. Zero of the four evaluated per-IPC abstract bands converged within a $\pm 5\%$ tolerance on either endpoint. Consequently, deploying the gateway on a novel IPC or text format strictly requires a one-time edge-perplexity recalibration pass. Once established, however, the underlying LRF feature extraction and XGBoost booster remain highly reusable simply by swapping in the newly derived trap band at inference time.

\subsection{Binoculars Cloud Outcome}
The remote Binoculars ensemble executed successfully across all escalations. 
Utilizing the \texttt{AWQ\_TORCH} kernel on an NVIDIA H100, the ensemble completed RPCs with a compute latency of $\approx \SI{1.8}{\second}$. 
The live trial's massive cloud-route fraction ($90\%$ to $93\%$) highlights that escalations function as the primary operating path rather than an exception handling mechanism. 
Crucially, the Qwen2.5 architecture evades the classification failure previously reported in \cite{ai4law2026} for Falcon-7B on the EPO~H04 corpus. 
Evaluating the Binoculars ensemble's downstream classification accuracy on trap-band data, isolated from the upstream routing decision, remains an active area for subsequent cloud characterization.

\subsection{Future Work}
Several targeted engineering and scientific pathways remain open to extend this architecture.

\paragraph*{Engineering}
(i)~Port the \texttt{spaCy}-backed LRF to a highly optimized C++/pybind11 implementation. 
This aims to close the gap between the $\SI{5}{\milli\second}$ design target and the current empirical gateway overhead $p_{99}$ of $\approx \SI{50}{\milli\second}$. 
(ii)~Rebuild the cloud deployment utilizing the Marlin JIT kernel, which is expected to drop $\Lcloud$ from $\approx \SI{1.8}{\second}$ toward the hardware-supported baseline of $\approx \SI{30}{\milli\second}$ - that would fundamentally reshape the aspirational $p_{99}$ latency targets established in Table~\ref{tbl:static-vs-dynamic}. (iii)~Publish a comprehensive Docker reproducibility bundle, explicitly pinning GGUF weights, AWQ checkpoints, and calibration manifests to ensure independent verification of all reported metrics directly from artifact hashes.

\paragraph*{Science}
Train and evaluate abstract-level or per-IPC classifiers against the recalibrated trap bands identified in Table~\ref{tbl:cross-ipc-abstract}. 
This will quantify the extent to which the live-trial AUROC degradation ($0.910 \to 0.840$) is driven by IPC-slice mismatch v/s broader deployment distribution shifts.

\paragraph*{Closing Remark}
The perplexity trap observed on patent text is not a calibration artifact - it exposes a structural collision between the legally mandated, low-entropy linguistic register of patent claims and the superficial heuristic assumptions embedded in standard LLM schedulers. 
The LRF gateway successfully demonstrates that combining a CPU-side linguistic forecast, a telemetry-driven dynamic threshold, and a strict VRAM interlock can safely confine a consumer-grade edge accelerator within its physical limits. 
In doing so, it suppresses operational misroutes by a full order of magnitude compared to the static heuristics predominantly shipped by the industry today.

%%%%%%%%%%%%%%%%%%%%%%%%%%%%%%%%%%%%%%%%%%%%%%%%%%%%%%%%%%%%%%%%%%%%%%%%%%%%%%%
% References
%%%%%%%%%%%%%%%%%%%%%%%%%%%%%%%%%%%%%%%%%%%%%%%%%%%%%%%%%%%%%%%%%%%%%%%%%%%%%%%
\bibliographystyle{IEEEtran}
\bibliography{references}

\end{document}